%% file: main.tex
\documentclass[letterpaper,journal]{IEEEtran}
\usepackage{amsmath,amssymb,amsfonts}
\usepackage{amsthm}

\theoremstyle{definition}
\newtheorem{definition}{Definition}

\usepackage{cite}

\usepackage{graphicx}
\usepackage[caption=false,font=normalsize,labelfont=sf,textfont=sf]{subfig}
\usepackage{tikz}
\usetikzlibrary{arrows.meta,calc,positioning}

\usepackage{booktabs}
\usepackage{multirow}
\usepackage{array}
\usepackage{tabularx}
\newcolumntype{Y}{>{\centering\arraybackslash}X}

\IfFileExists{algorithm.sty}{
\usepackage{algorithm}
}{
\usepackage{float}
\floatstyle{ruled}
\newfloat{algorithm}{!htb}{loa}
\floatname{algorithm}{Algorithm}
}
\newif\ifhasalgorithmic
\IfFileExists{algorithmic.sty}{\hasalgorithmictrue}{\hasalgorithmicfalse}
\ifhasalgorithmic
\usepackage{algorithmic}
\else
\newenvironment{algorithmic}[1][]%
{\begin{list}{}{\setlength{\leftmargin}{1em}\setlength{\itemsep}{0pt}\setlength{\parsep}{0pt}}}%
{\end{list}}
\newcommand{\REQUIRE}{\item[\textbf{Require:}]}

\newcommand{\STATE}{\item[]}
\newcommand{\WHILE}[1]{\item[\textbf{while} #1 \textbf{do}]}
\newcommand{\ENDWHILE}{\item[\textbf{end while}]}
\newcommand{\IF}[1]{\item[\textbf{if} #1 \textbf{then}]}
\newcommand{\ENDIF}{\item[\textbf{end if}]}
\fi

\usepackage{textcomp}
\usepackage{xcolor}
\usepackage{url}
\usepackage{verbatim}

\usepackage{stfloats}
\usepackage{placeins}

\usepackage[hidelinks]{hyperref}

\begin{document}
\flushbottom
\bstctlcite{BSTcontrol}

\title{GuidedRay: Diversity-Guided Direction Discovery for Targeted Hard-Label Black-Box Attacks}
%GuidedRay: A Highly Query-Efficient Decision-based Targeted Adversarial Attack Based on Diversity-Oriented Sampling

\author{
Fei~Yuan,
Yantian~Shen,
Qingyuan~Yu,
Yi~Chen,\\
Binghui~Wang,~\IEEEmembership{Senior Member,~IEEE},
Hongbo~Yu,
Anyu~Wang,
and~Xiaoyun~Wang%
\thanks{(Corresponding authors: Yi Chen and Binghui Wang.)}
\thanks{Fei Yuan is with Shandong University, Jinan, Shandong 250100, China (e-mail:
\nolinkurl{yuanfei@mail.sdu.edu.cn}).}
\thanks{Yantian Shen, Yi Chen, Hongbo Yu, Anyu Wang, and Xiaoyun Wang are with
Tsinghua University, Beijing 100084, China (e-mail:
\nolinkurl{shenyt22@mails.tsinghua.edu.cn};
\nolinkurl{chenyi2023@tsinghua.edu.cn};
\nolinkurl{yuhongbo@mail.tsinghua.edu.cn};
\nolinkurl{anyuwang@tsinghua.edu.cn};
\nolinkurl{xiaoyunwang@tsinghua.edu.cn}).}
\thanks{Qingyuan Yu is with the Institute of Satellite Information Engineering,
Beijing 100094, China (e-mail:
\nolinkurl{yu_qingyuan@yeah.net}).}
\thanks{Binghui Wang is with the Department of Computer Science, Illinois
Institute of Technology, Chicago, IL 60616 USA (e-mail:
\nolinkurl{bwang70@illinoistech.edu}).}
}

\maketitle

\begin{abstract}
Deep neural networks are vulnerable to adversarial attacks. Among black-box attacks, targeted decision-based attacks are particularly difficult: the attacker observes only the target model's top-1 label and aims to make it predict a prespecified target class under a bounded perturbation. Before perturbation refinement, the attacker must discover a direction that reaches the prescribed target region. This initialization step can incur substantial query cost. We propose GuidedRay, a targeted decision-based attack based on diversity-guided direction discovery. GuidedRay builds on two observations: target-class reference samples provide useful target-conditioned direction priors, and diverse candidates increase the probability of discovering a targeted adversarial direction. GuidedRay generates varied candidates from one or multiple target-class references and uses a one-query Fast Test to screen their induced sign directions. Once a feasible direction is found, GuidedRay applies Ray Search to reduce its decision-boundary radius. Experiments on CIFAR-10, CIFAR-100, and ImageNet demonstrate that GuidedRay consistently outperforms five state-of-the-art decision-based attacks at four evaluated query budgets from 500 to 5,000, with particularly pronounced gains in direction discovery during initialization. Against models protected by adversarial training or TRADES, it likewise achieves the highest attack success rate at all four query budgets.
\end{abstract}

\begin{IEEEkeywords}
Adversarial machine learning, black-box adversarial attacks, decision-based attacks, direction discovery, targeted adversarial attacks.
\end{IEEEkeywords}

\input{sections/introduction}

\input{sections/preliminaries_background}

\input{sections/problem_statement}

\input{sections/heuristic_motivations}

\input{sections/proposed_attack}

\input{sections/experiments}

\FloatBarrier
\section{Conclusion}
This paper studies targeted adversarial-direction discovery in decision-based
attacks under hard-label black-box access. Guided by target-class priors and
candidate diversity, GuidedRay uses ADSF to generate varied target-conditioned
directions and screen them with a one-query Fast Test before Ray Search
refinement. Experiments on three benchmark datasets show that GuidedRay
consistently outperforms the evaluated baselines across different query budgets
and remains effective against defended models. Ablation studies further
demonstrate the contribution of ADSF and GuidedRay's greater candidate-direction
diversity relative to RayS. Untargeted results demonstrate broader
applicability, with smaller gains under the less restrictive objective. These
findings highlight targeted-direction discovery as an important design focus
for decision-based attacks.

\bibliographystyle{IEEEtran}
\bibliography{references}

\end{document}

%% file: sections/introduction.tex
% !TeX spellcheck = en_US
% !TEX root = main.tex

\section{Introduction}
\label{sec:introduction}

\IEEEPARstart{D}{eep} neural networks (DNNs) are vulnerable to adversarial
attacks, in which carefully crafted perturbations cause a model to produce
incorrect predictions~\cite{DBLP:journals/corr/SzegedyZSBEGF13,
DBLP:journals/corr/GoodfellowSS14}. Adversarial attacks can be either
untargeted or targeted. An untargeted attack only requires the prediction to
differ from the ground-truth class, whereas a targeted attack must force the
model to output a specific class chosen by the attacker. Reaching a
pre-specified target class is substantially more restrictive than inducing
arbitrary misclassification.

Besides the attack objective, the information exposed by the victim model
defines the attack setting. This work considers targeted decision-based
attacks, which operate in a hard-label black-box setting. The attacker has no
access to the model architecture, parameters, gradients, logits, or confidence
scores and observes only the top-1 predicted label. Given a benign input, a
target class, a perturbation bound, and a query budget, the attacker aims to
construct an adversarial example that is classified as the target class. The
combination of a prespecified attack objective and label-only feedback makes
targeted decision-based attacks particularly difficult~\cite{
DBLP:journals/csur/LiXGYX24,DBLP:conf/iclr/BrendelRB18,
DBLP:conf/sp/ChenJW20}.

Many existing targeted decision-based attacks begin with a reference input that
is already classified as the target class and then iteratively reduce its
distance to the benign input. Existing studies have developed boundary estimation~\cite{
DBLP:conf/iclr/BrendelRB18,DBLP:conf/sp/ChenJW20}, direction
optimization~\cite{DBLP:conf/iclr/ChengLCZYH19,
DBLP:conf/iclr/ChengSCC0H20}, and geometric acceleration~\cite{
DBLP:conf/nips/MaGCYW21,DBLP:conf/eccv/WangZTGHLL22,
DBLP:conf/cvpr/MahoFM21,DBLP:conf/iccv/RezaR0D23,
DBLP:conf/sp/WanFWY24} to improve this refinement process. These methods
mainly address how to reduce the perturbation once an adversarial example or a
useful search direction has been obtained. Before refinement can proceed,
however, the attacker must first obtain a direction that reaches the prescribed
target region.

Under the constraint of the $L_\infty$ norm, adversarial perturbations obtained
from two powerful white-box attacks,
FGSM~\cite{DBLP:journals/corr/GoodfellowSS14} and
PGD~\cite{DBLP:conf/iclr/MadryMSTV18}, are often found on the vertices of the
$L_\infty$ norm ball~\cite{DBLP:conf/icml/MoonAS19,
DBLP:conf/aaai/ChenZYG20}. Motivated by this observation, our work restricts
the search space to vertex-pointing rays parameterized by sign directions and
represents a perturbed point as $\mathbf{x}_b+r\cdot\mathbf{d}$, where
$\mathbf{x}_b$ denotes a benign input, $\mathbf{d}\in\{-1,+1\}^n$ is a sign
direction, and $r>0$ is the radius. Each sign direction $\mathbf{d}$ defines a
ray starting from $\mathbf{x}_b$. More specifically, along the ray induced by
$\mathbf{d}$, the smallest $r$ for which $\mathbf{x}_b+r\cdot\mathbf{d}$ is
classified as the prescribed target class is called the
\emph{decision-boundary radius}. This representation naturally
decomposes an attack into two stages: initialization discovers a sign direction
along which the target region can be reached, whereas optimization refines that
direction to reduce its decision-boundary radius.
RayS~\cite{DBLP:conf/kdd/ChenG20} demonstrates that systematic search in the
discrete sign space $\{-1,+1\}^n$ can be effective for the untargeted
objective. When evaluated in our targeted setting, however, RayS does not
consistently improve over existing targeted attacks across the evaluated
datasets and query budgets, as shown in Table~\ref{tab:ASR-standard} of
Section~\ref{sec:experiments}. Unlike an untargeted attack, targeted
initialization must locate a direction that reaches one particular target
region rather than any incorrect class, making the discovery of targeted
adversarial directions during initialization much more difficult.

The differing requirements of untargeted and targeted initialization motivate
a broader examination of targeted sign-space initialization. Two factors
contribute to this difficulty. First, a direction sampled without target
information is unlikely to reach a prespecified target region in a
high-dimensional sign space, especially for classification tasks with many
classes. Second, although a target-class reference sample $\mathbf{x}_a$
provides useful information about the target region, this information must be
translated into a discrete sign direction $\mathbf{d}\in\{-1,+1\}^n$.
Following the discrete representation used by
RayS~\cite{DBLP:conf/kdd/ChenG20}, we map the displacement from the benign
sample $\mathbf{x}_b$ toward $\mathbf{x}_a$ to the sign direction
$\mathbf{d}_a=\operatorname{sign}(\mathbf{x}_a-\mathbf{x}_b)$. Because this
mapping retains only the coordinate-wise sign information, the induced
direction is not necessarily feasible.

We investigate this targeted-direction discovery challenge through two
observations. First,
target-class reference samples provide a useful target-conditioned prior:
their induced directions reach the target region more often than directions
induced by non-target references. Second, repeatedly testing highly similar
directions explores nearby parts of the sign space and provides limited
additional coverage, whereas a diverse candidate set is more likely to contain
a targeted adversarial direction. The motivating experiments in
Section~\ref{sec:heuristic_motivations} provide empirical evidence for these
observations and guide the design of a direction-discovery strategy that
combines target-class guidance with candidate diversity.

Guided by these observations, we propose GuidedRay, which consists of an
initialization phase and an optimization phase. During initialization, we
develop an \emph{Adversarial Direction Search Framework} (ADSF) to discover a
targeted adversarial direction. ADSF uses diversity-oriented sample augmentation
to generate varied candidates from one or multiple target-class reference
samples and maps them to sign directions relative to the benign input. It then
uses a one-query Fast Test to screen each direction without performing an
expensive boundary search for every candidate. Once ADSF finds a targeted
adversarial direction, the optimization phase applies Ray Search~\cite{
DBLP:conf/kdd/ChenG20} to reduce its decision-boundary radius. In this way,
ADSF addresses the targeted-direction discovery problem, while Ray Search
performs the subsequent direction optimization in the same discrete space.

We compare GuidedRay with five representative decision-based attacks on
CIFAR-10, CIFAR-100, and ImageNet. Across the three datasets, GuidedRay obtains
the highest attack success rate (ASR) at all four primary query budgets:
500, 1,000, 3,000, and 5,000. Section~\ref{subsubsec:diversity} further shows
that GuidedRay generates more diverse directions, with higher initialization
success rates and fewer initialization queries than RayS. Against models
protected by adversarial training~\cite{DBLP:conf/iclr/MadryMSTV18} or
TRADES~\cite{DBLP:conf/icml/ZhangYJXGJ19}, GuidedRay also achieves the highest
ASR at all four query budgets. These results demonstrate that
strengthening targeted-direction discovery can move successful attacks to
earlier query budgets.

\noindent\textbf{Our Contributions.}
The main contributions of this work are summarized as follows:
\begin{itemize}
\item We identify the discovery of targeted adversarial directions in the
discrete sign space as a major bottleneck in targeted decision-based attacks under the
hard-label black-box setting, and further conduct motivating experiments
showing that target-class guidance provides an informative initialization
prior and that candidate diversity broadens sign-space coverage, both
facilitating feasible-direction discovery.

\item We develop ADSF, whose candidate-generation module combines target-class
guidance with diversity-oriented sample augmentation. A one-query Fast Test
then screens the induced directions. Based on ADSF, GuidedRay discovers an
adversarial direction during initialization and then refines its decision-boundary
radius using Ray Search.

\item We conduct extensive experiments comparing GuidedRay with five decision-based attacks on
three benchmark datasets and adversarially trained models. The results
demonstrate the consistent effectiveness of GuidedRay across different
datasets, query budgets, and model settings. Ablation studies further verify
the contribution of ADSF and show that GuidedRay generates more diverse
candidate directions than RayS while achieving higher ISR and lower
initialization query consumption.
\end{itemize}

Code for all experiments is available at
\url{https://github.com/sudyuan/GuidedRay}.

%% file: sections/preliminaries_background.tex
% !TeX spellcheck = en_US
% !TEX root = main.tex

\section{Related Work}
\label{sec:background}
Adversarial attacks can be broadly categorized into white-box and black-box
attacks according to the information available to the adversary. In white-box
attack research, the primary focus has been on finding adversarial examples
with minimal perturbation magnitude~\cite{
DBLP:journals/corr/SzegedyZSBEGF13,
DBLP:journals/corr/GoodfellowSS14,
DBLP:conf/iclr/KurakinGB17a,
DBLP:conf/iclr/MadryMSTV18,
DBLP:conf/sp/Carlini017,
DBLP:conf/eurosp/PapernotMJFCS16,
DBLP:conf/pkdd/BiggioCMNSLGR13,
DBLP:conf/cvpr/Moosavi-Dezfooli16}. In contrast, research on black-box attacks
has centered on efficiently generating adversarial examples, with existing
methods generally falling into three categories: transfer-based attacks~\cite{
DBLP:conf/ccs/PapernotMGJCS17,
DBLP:journals/corr/PapernotMG16,
DBLP:conf/iclr/LiuCLS17,
DBLP:conf/cvpr/DongLPS0HL18, 
DBLP:conf/cvpr/DongPSZ19}, 
score-based attacks~\cite{
DBLP:conf/ccs/ChenZSYH17,
DBLP:conf/iclr/IlyasEM19,
DBLP:conf/icml/IlyasEAL18,
DBLP:conf/gecco/AlzantotSCZHS19}, 
and decision-based attacks. This work focuses on decision-based attacks.

Boundary Attack~\cite{DBLP:conf/iclr/BrendelRB18} introduced the
decision-based setting, in which the adversary observes only the top-1
predicted label. Most subsequent methods start from an input satisfying the
attack objective and improve it through boundary estimation, direction
optimization, or geometric search. HSJA~\cite{DBLP:conf/sp/ChenJW20} and
AHA~\cite{DBLP:conf/iccv/LiJC0HZLLHW21} estimate local boundary information
from hard-label queries; OPT~\cite{DBLP:conf/iclr/ChengLCZYH19} and
Sign-OPT~\cite{DBLP:conf/iclr/ChengSCC0H20} optimize the search direction based
on perturbation changes; and Tangent~\cite{DBLP:conf/nips/MaGCYW21},
Triangle~\cite{DBLP:conf/eccv/WangZTGHLL22},
SurFree~\cite{DBLP:conf/cvpr/MahoFM21},
CGBA~\cite{DBLP:conf/iccv/RezaR0D23}, and TtBA~\cite{Wang2025TtBA} exploit
geometric properties of the decision boundary to accelerate refinement. These
methods primarily focus on reducing perturbation after a useful adversarial
point or direction has been obtained.

Some recent hard-label attacks introduce additional attacker resources.
SQBA~\cite{Park2024SQBA}, DEAL~\cite{DBLP:journals/tdsc/ShenLYLZX24}, and
Prior-OPT/Prior-Sign-OPT~\cite{Ma2025PriorRay} use a pretrained surrogate
model, an offline meta-learned boundary learner, and surrogate-gradient priors,
respectively. GuidedRay assumes none of these auxiliary models and operates
only with hard-label queries and attacker-available target-class references.

Distinct from the continuous-space refinement strategies above,
RayS~\cite{DBLP:conf/kdd/ChenG20} formulates $L_\infty$ hard-label attacks as
a discrete search over sign directions. Its Ray Search procedure hierarchically
flips blocks of a sign direction and retains changes that reduce the
decision-boundary radius. GuidedRay adopts the same discrete representation and
uses Ray Search for subsequent optimization, but focuses on a different
bottleneck: discovering a feasible targeted direction during initialization.
It addresses this problem through target-class guidance and diverse
candidate-direction generation before refinement begins.

%% file: sections/problem_statement.tex
% !TeX spellcheck = en_US
% !TEX root = main.tex

\section{Threat Model and Problem Formulation}
\label{sec:threat_model}

\subsection{Threat Model}
We consider decision-based attacks against image classifiers under hard-label
black-box access. For each query, the victim returns only the top-1 predicted
label; the attacker cannot access the model architecture, parameters,
gradients, logits, confidence scores, or training data. Deploying a defense
does not change this access model.

Let the victim classifier be
$f:\mathbb{R}^n\rightarrow\{1,\ldots,L\}$, where $n$ is the input dimension
and $L$ is the number of classes. Let $\mathbf{x}_b\in\mathbb{R}^n$ be a correctly
classified benign input and denote its predicted class by
$y_b=f(\mathbf{x}_b)$. The attacker is given a preset target class
$t\ne y_b$, a perturbation threshold $\epsilon>0$, and a maximum query budget
$Q$. The attacker's goal is to construct an adversarial example classified as
$t$ while satisfying the perturbation constraint within $Q$ queries.

The attacker also has access to $k\geq1$ correctly classified target-class
reference samples $\{\mathbf{x}_a^{(j)}\}_{j=1}^{k}$ satisfying
$f(\mathbf{x}_a^{(j)})=t$. These samples provide information about the target
class but are not required to satisfy the perturbation constraint relative to
$\mathbf{x}_b$. We call $k=1$ and $k>1$ the \emph{single-reference} and
\emph{multi-reference} settings, respectively. Their availability does not
imply access to the victim's training data, and GuidedRay uses neither a
substitute model nor a local boundary learner.

\subsection{Definitions and Notation}
\label{subsec:defs}
\begin{definition}[Adversarial region]
The adversarial region $O$ is defined as
\begin{equation}
\label{eq:adversarial_space}
O=
\begin{cases}
\mathbf{x}\in\mathbb{R}^n:f(\mathbf{x})\ne f(\mathbf{x}_b),
& \text{untargeted attack},\\
\mathbf{x}\in\mathbb{R}^n:f(\mathbf{x})=t,
& \text{targeted attack}.
\end{cases}
\end{equation}
For targeted attacks, we denote the target-class region
$\{\mathbf{x}\in\mathbb{R}^n:f(\mathbf{x})=t\}$ by $A_t$.
\end{definition}

\begin{definition}[Adversarial example]
\label{adv_example}
An input $\mathbf{x}_{\mathrm{adv}}$ is an adversarial example for
$\mathbf{x}_b$ if
$\mathbf{x}_{\mathrm{adv}}\in O$ and
$\|\mathbf{x}_{\mathrm{adv}}-\mathbf{x}_b\|_\infty\leq\epsilon$.
\end{definition}

Following the discrete representation introduced in
Section~\ref{sec:introduction}, GuidedRay restricts its search to sign
directions $\mathbf{d}\in\{-1,+1\}^n$ and considers points of the form
$\mathbf{x}_b+r\mathbf{d}$, where $r>0$ is the radius.

\begin{definition}[Adversarial direction]
\label{adv_direction}
A sign direction $\mathbf{d}\in\{-1,+1\}^n$ is an adversarial direction if
there exists $r>0$ such that $\mathbf{x}_b+r\mathbf{d}\in O$.
\end{definition}
For targeted attacks, where $O=A_t$, we refer to such a direction as a
\emph{targeted adversarial direction}. A target-class reference sample
$\mathbf{x}_a^{(j)}\in A_t$ induces the sign direction
$\mathbf{d}_a^{(j)}
=\operatorname{sign}(\mathbf{x}_a^{(j)}-\mathbf{x}_b)$, where
$\operatorname{sign}(\cdot)$ is applied coordinate-wise and zero-valued
entries are mapped to $+1$.

\subsection{Problem Formulation}
Unless otherwise specified, the following discussion focuses on targeted
attacks, for which $O=A_t$. For a sign direction $\mathbf{d}$, its
decision-boundary radius is defined as
\begin{equation}
g(\mathbf{d})
=
\inf\left\{
r>0\ \middle|\ \mathbf{x}_b+r\mathbf{d}\in O
\right\}.
\label{eq:decision_boundary_radius}
\end{equation}
If the set is empty, $g(\mathbf{d})=+\infty$; otherwise, $\mathbf{d}$ is an
adversarial direction and $g(\mathbf{d})$ is the smallest radius at which its
ray reaches the adversarial region.

Subject to the query budget, the attack seeks a sign direction with the
smallest attainable decision-boundary radius:
\begin{equation}
\min_{\mathbf{d}\in\{-1,+1\}^n} g(\mathbf{d})
\qquad
\text{subject to}\quad q\leq Q,
\label{eq:attack_objective}
\end{equation}
where $q$ denotes the number of model queries used by the attack. An attack
run is successful if its returned direction $\widehat{\mathbf{d}}$ satisfies
$g(\widehat{\mathbf{d}})\leq\epsilon$.

Initialization finds a direction with $g(\mathbf{d})<+\infty$, and optimization
reduces its radius under the remaining budget. GuidedRay addresses the former
and uses Ray Search for the latter. The formulation also applies to untargeted
attacks using the first case of Eq.~\eqref{eq:adversarial_space}; the
corresponding GuidedRay variant is presented in
Section~\ref{subsec:untargeted_variant}.

%% file: sections/heuristic_motivations.tex
% !TeX spellcheck = en_US
% !TEX root = main.tex

\section{Heuristic Motivation for Targeted Direction Discovery}
\label{sec:heuristic_motivations}

Existing hard-label black-box attacks mainly refine perturbations after an
adversarial direction has been found. In targeted attacks, however,
discovering such an initial direction is itself a major bottleneck. Since the
attacker can only observe the top-1 predicted label and must reach a
pre-specified target region, targeted adversarial directions may be sparse in
the discrete sign space $\{-1,+1\}^n$.

Motivated by this observation, we revisit the initialization stage from two
heuristic perspectives. First, target-class reference samples may provide
useful target-conditioned priors for discovering adversarial directions.
Second, candidate directions should be sufficiently diverse, as highly similar
directions tend to explore nearby regions of the sign space and provide limited
additional coverage.

\subsection{Motivation 1: Target-Conditioned Direction Prior}
\label{subsec:motivation_target_prior}

The first motivation concerns how to obtain an informative initialization prior
when the attacker can observe only hard labels. Using the notation introduced
in Section~\ref{sec:threat_model}, a sign direction
$\mathbf{d}\in\{-1,+1\}^n$ is a targeted adversarial direction if there
exists $r>0$ such that $\mathbf{x}_b+r\mathbf{d}\in A_t$. Randomly sampling
from the sign space ignores the identity of the prescribed target class and
therefore provides no target-specific guidance for finding such a direction.

A target-class reference $\mathbf{x}_a\in A_t$ provides an observed point in
the prescribed target region. Its displacement from the benign input contains
coordinate-wise information about how that target point differs from
$\mathbf{x}_b$. Mapping the displacement to
$\mathbf{d}_a=\operatorname{sign}(\mathbf{x}_a-\mathbf{x}_b)$ retains this
coordinate-wise orientation while conforming to the discrete search space in
Section~\ref{subsec:defs}. As the sign operation discards relative
coordinate magnitudes, $\mathbf{x}_a$ does not generally equal
$\mathbf{x}_b+r\mathbf{d}_a$ for any scalar $r$, and the induced sign
direction is not guaranteed to reach $A_t$.

Nevertheless, unlike a direction induced by an unrelated reference,
$\mathbf{d}_a$ is constructed from an observed target-class point and thus
provides a target-conditioned initialization prior. This motivates the
heuristic that target-class references are more likely than non-target
references to induce adversarial directions. We evaluate this hypothesis
empirically below.

\subsubsection{Motivating experiments}
To evaluate this hypothesis empirically, we compare the feasibility of
directions induced by target-class and non-target references on CIFAR-10/ResNet-50,
CIFAR-100/ResNet-50, and
ImageNet/DenseNet-121. For each dataset, we sample 5,000 correctly classified
benign inputs. Given a benign input $\mathbf{x}_b$ and a randomly selected
target class $t$, \textsc{Random-Reference} constructs a sign direction from a
randomly selected non-target reference sample $\mathbf{x}_r\notin A_t$, while
\textsc{Target-Reference} constructs a sign direction from a target-class
reference sample $\mathbf{x}_a\in A_t$. For each direction $\mathbf{d}$, we
scan points of the form $\operatorname{clip}(\mathbf{x}_b+r\mathbf{d},0,1)$
over 200 radii in $[0,1]$ and regard $\mathbf{d}$ as successful if any
queried point is classified as the target class $t$.

Table~\ref{tab:target_prior_validation} reports the success rate, defined as
the fraction of sampled directions that reach $A_t$ along the scanned ray.
Across all three datasets, directions induced by target-class references
achieve higher success rates than those induced by random non-target
references. This result supports the heuristic that target-class references
provide useful target-conditioned priors for discovering adversarial
directions.

\begin{table}[H]
\centering
\renewcommand{\arraystretch}{1.05}
\setlength{\tabcolsep}{3pt}
\caption{Success rates of directions induced by random and target-class
references.}
\label{tab:target_prior_validation}
\begin{tabular*}{\columnwidth}{@{\extracolsep{\fill}}lcc@{}}
\toprule
Dataset / Model & \shortstack{Random-reference\\SR $\uparrow$} &
\shortstack{Target-reference\\SR $\uparrow$} \\
\midrule
\shortstack{CIFAR-10 /\ ResNet-50} & 0.2148 & 0.3866 \\
\shortstack{CIFAR-100 /\ ResNet-50} & 0.0360 & 0.1186 \\
\shortstack{ImageNet /\ DenseNet-121} & 0.0052 & 0.1012 \\
\bottomrule
\end{tabular*}
\end{table}

\subsection{Motivation 2: Diverse Candidate Directions}
\label{subsec:motivation_low_redundancy}

Motivation~1 suggests that target-class reference samples provide useful
target-conditioned priors for discovering adversarial directions.
However, generating multiple target-conditioned directions does not
necessarily improve initialization if they are highly similar, because such
directions repeatedly explore nearby parts of the discrete sign space.
Candidate diversity can therefore broaden sign-space coverage while retaining
target-class guidance.

For a candidate set
$\mathcal{D}=\{\mathbf{d}_1,\ldots,\mathbf{d}_m\}$ with
$\mathbf{d}_i\in\{-1,+1\}^n$, we use the average pairwise cosine similarity
(APCS) to measure directional redundancy:
\begin{equation}
\mathrm{APCS}(\mathcal{D})
=
\frac{2}{m(m-1)}
\sum_{1\le i<j\le m}
\textrm{cos}(\mathbf{d}_i,\mathbf{d}_j),
\label{eq:apcs_motivation}
\end{equation}
where $\textrm{cos}(\cdot,\cdot)$ is the cosine similarity. A smaller APCS
indicates lower directional redundancy and broader exploration of the sign
space. However, diversity should be
considered together with target-class guidance, because increasing diversity
indiscriminately may discard the useful prior identified in Motivation~1.

\subsubsection{Motivating experiments}
To examine the effect of candidate diversity, we conduct a candidate-direction
analysis on CIFAR-10/ResNet-50. For each trial, we generate $m=200$ candidate
directions and evaluate their feasibility using the same ray-scanning procedure
as in Motivation~1. Results are averaged over 100 trials. \textsc{RayS} is
included as a sign-space reference; since it does not use a target-class
reference, the sign change ratio defined below is not applicable to it.

Starting from a target-conditioned seed direction
$\mathbf{d}_a=\operatorname{sign}(\mathbf{x}_a-\mathbf{x}_b)$ with
$\mathbf{x}_a\in A_t$, we construct a candidate pool by adding Gaussian noise
to $\mathbf{x}_a$, mixing it with another target-class reference, or sampling
another target-class reference. For this analysis, we retain
samples $\mathbf{x}'$ satisfying $f(\mathbf{x}')=t$ and form directions
$\operatorname{sign}(\mathbf{x}'-\mathbf{x}_b)$. For each retained candidate,
the sign change ratio (SCR) relative to $\mathbf{d}_a$ is
\[
\mathrm{SCR}(\mathbf{x}')
=
\frac{\mathrm{Ham}(\operatorname{sign}(\mathbf{x}'-\mathbf{x}_b),\mathbf{d}_a)}{n}.
\]
The low-, medium-, and high-SCR groups comprise the bottom 20\%, 40th--60th
percentiles, and top 20\% of the retained candidates, respectively. The
label-based filtering and exhaustive ray scans are used only to
construct and evaluate these SCR-stratified groups; they are not steps of
GuidedRay.

We report SCR, APCS, direction success rate (DSR), and candidate-set success
rate (CSSR). DSR is the fraction of valid targeted adversarial
directions, whereas CSSR is the probability that a candidate set contains at
least one such direction. Table~\ref{tab:low_redundancy_validation} reports the
detailed CIFAR-10/ResNet-50 results, while
Table~\ref{tab:motivation_cross_dataset_summary} summarizes CSSR across all
three datasets.

\begin{table}[H]
\centering
\renewcommand{\arraystretch}{1.05}
\setlength{\tabcolsep}{2pt}
\caption{SCR-stratified candidate-direction diversity on
CIFAR-10/ResNet-50.}
\label{tab:low_redundancy_validation}
\begin{tabular*}{\columnwidth}{@{\extracolsep{\fill}}lcccc@{}}
\toprule
Method / Level & SCR & APCS $\downarrow$ & DSR $\uparrow$ & CSSR $\uparrow$ \\
\midrule
\textsc{RayS} & -- & 0.8536 & 0.1855 & 0.7200 \\
Low SCR    & 0.0306 & 0.9129 & 0.3810 & 0.7300 \\
Medium SCR & 0.2584 & 0.4978 & 0.3944 & 0.9200 \\
High SCR   & 0.4271 & 0.3229 & 0.4032 & 0.9800 \\
\bottomrule
\end{tabular*}
\end{table}

\begin{table}[H]
\centering
\renewcommand{\arraystretch}{1.05}
\setlength{\tabcolsep}{2pt}
\caption{Cross-dataset results for Motivation~2, averaged over 100
trials with $m=200$ directions per trial.}
\label{tab:motivation_cross_dataset_summary}
\begin{tabular*}{\columnwidth}{@{\extracolsep{\fill}}lcc@{}}
\toprule
Dataset / Model & \shortstack{\textsc{RayS}\\CSSR $\uparrow$} &
\shortstack{High-SCR\\CSSR $\uparrow$} \\
\midrule
\shortstack{CIFAR-10 /\\ResNet-50} & 0.7200 & 0.9800 \\
\shortstack{CIFAR-100 /\\ResNet-50} & 0.3800 & 0.5000 \\
\shortstack{ImageNet /\\DenseNet-121} & 0.0600 & 0.5500 \\
\bottomrule
\end{tabular*}
\end{table}

Across the SCR-stratified groups, increasing SCR reduces APCS, indicating
lower directional redundancy. More importantly, CSSR
increases from 0.7300 in the low-SCR group to 0.9200 and 0.9800 in the
medium- and high-SCR groups, respectively. Thus, the more diverse candidate
sets are more likely to contain an adversarial direction. Although
\textsc{RayS} provides a useful sign-space reference, it differs from the
SCR-stratified groups in both its generation mechanism and its use of
target-class information.

Overall, these results support the second heuristic motivation: among
target-conditioned candidates, greater diversity corresponds to lower
directional redundancy, broader sign-space coverage, and a higher observed
probability that a candidate set contains an adversarial direction.

%% file: sections/proposed_attack.tex
% !TeX spellcheck = en_US
% !TEX root = main.tex

\section{GuidedRay}
\label{sec:guidedRay-overview}

\subsection{Overview}
\begin{figure*}[!t]
\centering
\begin{tikzpicture}
\node[inner sep=0] (overview) {\includegraphics[width=0.90\textwidth]{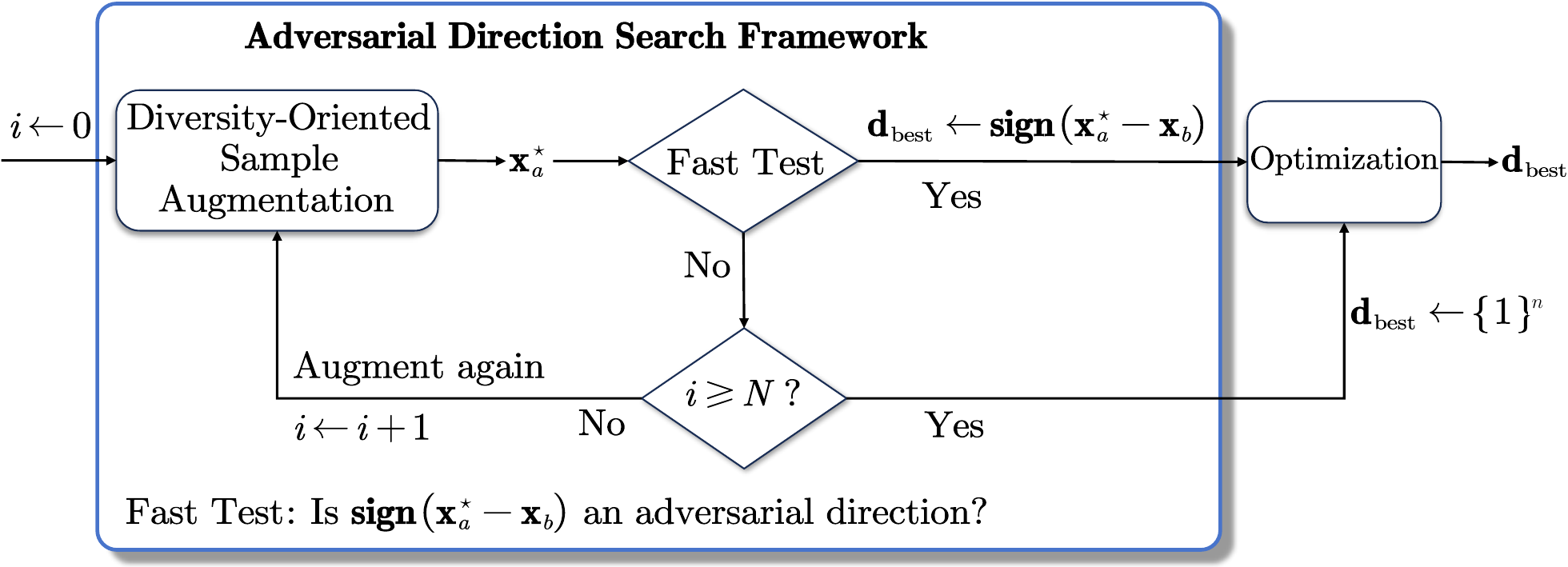}};
\node[anchor=south,fill=white,inner sep=1pt,font=\footnotesize\bfseries]
at ([xshift=-14mm,yshift=1mm]overview.north) {Initialization};
\end{tikzpicture}
\caption{Overview of GuidedRay. ADSF generates target-conditioned candidates
and uses one victim-model query to screen each induced sign direction. After an
adversarial direction is accepted, Ray Search refines its decision-boundary
radius using the remaining query budget.}
\label{fig:overview_of_attack}
\end{figure*}

GuidedRay consists of an initialization phase and an optimization phase, as
illustrated in Fig.~\ref{fig:overview_of_attack}. Given a benign input
$\mathbf{x}_b$ and target-class references, the \emph{Adversarial Direction
Search Framework} (ADSF) generates varied candidates, maps them to sign
directions, and screens each direction using a one-query Fast Test. The
process continues until an adversarial direction is found or $N$ candidates
have been tested. If no direction is accepted, GuidedRay falls back to the
RayS all-ones direction $\mathbf{d}_{\rm best}=\{1\}^n$; otherwise, Ray Search
refines the accepted direction under the remaining query budget.

\subsection{Adversarial Direction Search Framework}
\label{sec:design_details}
ADSF combines target-guided, diversity-oriented candidate generation with a
one-query Fast Test that screens each induced direction.

\subsubsection{Candidate Generation}
\label{subsec:AEA}
Let $\{\mathcal{T}_1,\ldots,\mathcal{T}_m\}$ denote a collection of image
augmentation methods. At each initialization step, ADSF selects a target-class
reference $\mathbf{x}_a^{(j)}$ and an augmentation method $\mathcal{T}$, and
generates a candidate
$\mathbf{x}_a^\star=\mathcal{T}(\mathbf{x}_a^{(j)})$. It then maps this candidate to
the discrete sign space through
$\mathbf{d}_c=\operatorname{sign}(\mathbf{x}_a^\star-\mathbf{x}_b)$.
Varying the reference and augmentation method reduces redundancy among the
induced target-conditioned directions, consistent with the observations in
Section~\ref{sec:heuristic_motivations}.

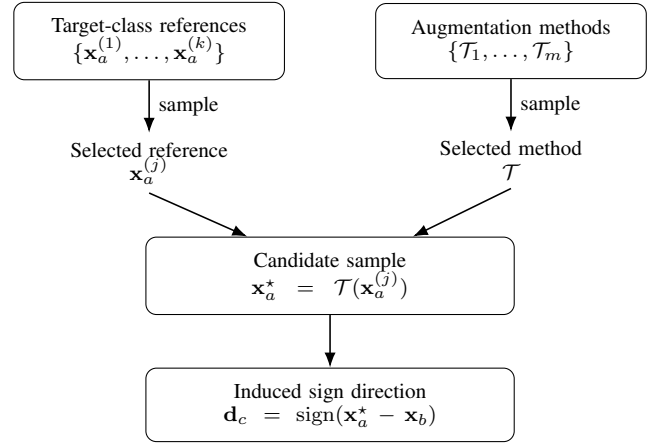
\begin{figure}[!t]
\centering
\resizebox{0.96\columnwidth}{!}{%
\begin{tikzpicture}[
    node distance=7mm and 12mm,
    >=Latex,
    every node/.style={font=\footnotesize,align=center},
    source/.style={draw,rounded corners,minimum height=10mm,
                   text width=32mm,inner sep=2mm},
    item/.style={minimum height=6mm,text width=26mm},
    result/.style={draw,rounded corners,minimum height=9mm,
                   text width=45mm,inner sep=2mm},
    flow/.style={->,semithick}
]
\node[source] (refs) {Target-class references\\
    $\{\mathbf{x}_a^{(1)},\ldots,\mathbf{x}_a^{(k)}\}$};
\node[source,right=of refs] (transforms) {Augmentation methods\\
    $\{\mathcal{T}_1,\ldots,\mathcal{T}_m\}$};
\node[item,below=of refs] (ref) {Selected reference\\$\mathbf{x}_a^{(j)}$};
\node[item,below=of transforms] (transform) {Selected method\\$\mathcal{T}$};
\node[result,below=10mm of $(ref)!0.5!(transform)$] (candidate)
    {Candidate sample\\
     $\mathbf{x}_a^\star=\mathcal{T}(\mathbf{x}_a^{(j)})$};
\node[result,below=of candidate] (direction)
    {Induced sign direction\\
     $\mathbf{d}_c=\operatorname{sign}(\mathbf{x}_a^\star-\mathbf{x}_b)$};

\draw[flow] (refs) -- node[right] {sample} (ref);
\draw[flow] (transforms) -- node[right] {sample} (transform);
\draw[flow] (ref.south) -- ([xshift=-11mm]candidate.north);
\draw[flow] (transform.south) -- ([xshift=11mm]candidate.north);
\draw[flow] (candidate) -- (direction);
\end{tikzpicture}
}
\caption{Target-conditioned candidate generation in ADSF. A reference and an
augmentation method are sampled to construct a candidate, which is then
mapped to a sign direction relative to the benign input.}
\label{fig:img_augmentation}
\end{figure}

For candidate generation, we randomly apply one of four augmentation methods
to each target-class reference: Gaussian noise, random rotation, random
cropping followed by resizing, or color jitter.
Fig.~\ref{fig:img_augmentation} illustrates the process.

\subsubsection{Single- and Multi-Reference Settings}
\label{subsubsec:k_AEGS}
In the single-reference setting ($k=1$), every candidate is generated from the
same target-class reference. In the multi-reference setting ($k>1$), ADSF
first samples one of the $k$ references and then applies a randomly selected
augmentation method. Using multiple references introduces additional
target-conditioned starting points and can broaden the coverage of the induced
directions. The effect of $k$ is evaluated in
Section~\ref{subsubsec:k_samples}.

\subsubsection{One-Query Fast Test}
\label{subsec:fast_test}
\begin{figure}[!t]
\centering
\includegraphics[width=0.62\columnwidth]{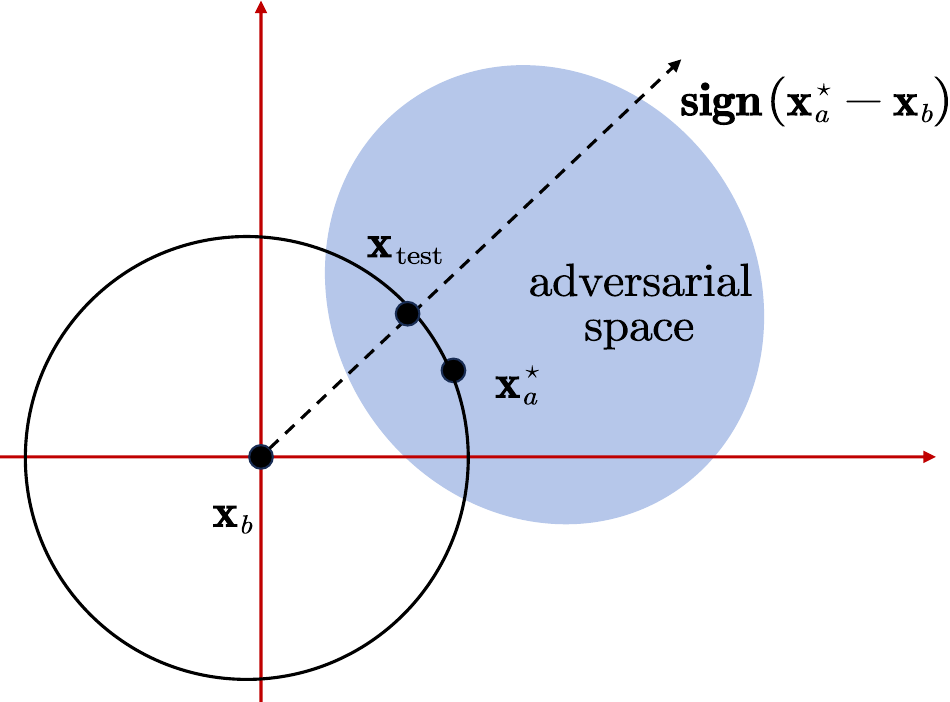}
\caption{Geometric illustration of a successful Fast Test. The shaded area
denotes the target adversarial region. If the queried point
$\mathbf{x}_{\rm test}$ lies in this region, the sign direction induced by
$\mathbf{x}_a^\star$ is verified as feasible.}
\label{fig:fast_test_geometry}
\end{figure}

Testing every candidate direction with a complete boundary search would
consume many queries before refinement begins. The Fast Test instead evaluates
one point on each induced ray. For a candidate $\mathbf{x}_a^\star$, let
$\mathbf{d}_c=\operatorname{sign}(\mathbf{x}_a^\star-\mathbf{x}_b)$ and
$\rho_c=\|\mathbf{x}_a^\star-\mathbf{x}_b\|_2$. The queried point is
\begin{equation}
\mathbf{x}_{\rm test}
=
\mathbf{x}_b
+
\rho_c
\frac{\mathbf{d}_c}{\|\mathbf{d}_c\|_2}.
\label{eq:fast_test}
\end{equation}
Fig.~\ref{fig:fast_test_geometry} illustrates a successful Fast Test, in
which the queried point lies in the target adversarial region.

If $f(\mathbf{x}_{\rm test})=t$, then $\mathbf{x}_{\rm test}\in A_t$ lies on
the ray induced by $\mathbf{d}_c$, at radius
$r_c=\rho_c/\|\mathbf{d}_c\|_2$ under the representation in
Section~\ref{subsec:defs}. The test therefore certifies
$\mathbf{d}_c$ as a targeted adversarial direction using one query.
GuidedRay accepts it as $\mathbf{d}_{\rm best}$ and performs a binary search
along the ray to obtain the initial radius $r_{\rm best}$.

If $f(\mathbf{x}_{\rm test})\neq t$, GuidedRay discards the candidate direction
without further queries, although it may be feasible at another radius.
Thus, ADSF accepts only verified adversarial directions while avoiding costly
boundary searches for candidates that fail the Fast Test.
Algorithm~\ref{alg:adsf} summarizes candidate generation, the Fast Test, and
the fallback direction.

\begin{algorithm}[!t]
\caption{ADSF Initialization}
\label{alg:adsf}
\begin{algorithmic}[1]
\REQUIRE
    Victim model $f$; benign input $\mathbf{x}_b$; target class $t$;\\
    references $\{\mathbf{x}_a^{(j)}\}_{j=1}^{k}$;
    augmentation methods $\{\mathcal{T}_i\}_{i=1}^{m}$;\\
    maximum number of initialization tests $N$.
\FOR{$i=1$ to $N$}
\STATE Sample a reference $\mathbf{x}_a^{(j)}$ and an augmentation method
       $\mathcal{T}$;
\STATE $\mathbf{x}_a^\star=\mathcal{T}(\mathbf{x}_a^{(j)})$;
\STATE $\mathbf{d}_c=\operatorname{sign}(\mathbf{x}_a^\star-\mathbf{x}_b)$,
       $\rho_c=\|\mathbf{x}_a^\star-\mathbf{x}_b\|_2$, and
       $r_c=\rho_c/\|\mathbf{d}_c\|_2$;
\STATE $y_{\rm test}=f(\mathbf{x}_{\rm test})$, with
       $\mathbf{x}_{\rm test}$ given by~\eqref{eq:fast_test};
\IF{$y_{\rm test}=t$}
\STATE $\mathbf{d}_{\rm best}=\mathbf{d}_c$;
\STATE $r_{\rm best}=\operatorname{SDBR}
       (f,\mathbf{x}_b,\mathbf{d}_{\rm best},r_c)$;
\STATE \textbf{return} $\mathbf{d}_{\rm best}$ and $r_{\rm best}$;
\ENDIF
\ENDFOR
\STATE \textbf{return} $\mathbf{d}_{\rm best}=\{1\}^n$ and
       $r_{\rm best}=+\infty$.
\end{algorithmic}
\end{algorithm}

\subsection{Optimization}
GuidedRay adopts Ray Search~\cite{DBLP:conf/kdd/ChenG20} because it refines the
decision-boundary radius in the same discrete sign space. Our design instead
focuses on targeted initialization. The ablations in
Sections~\ref{subsubsec:impact_of_aeaf} and~\ref{subsubsec:replace_ray_search}
further examine the contribution of ADSF independently of the subsequent
refinement procedure.

At search stage $s$, Ray Search divides the current sign direction into
$b_s=\min(2^s,n)$ blocks and flips one block at a time. A binary-search-based
boundary search, denoted by $\operatorname{SDBR}$, evaluates the temporary
direction. The current direction and radius are updated whenever the temporary
direction yields a smaller radius. The block granularity is then progressively
increased until the query budget is exhausted or the perturbation threshold is
met. Algorithm~\ref{alg:ray_search} summarizes this refinement procedure.

\begin{algorithm}[H]
\caption{Ray Search~\cite{DBLP:conf/kdd/ChenG20}}
\label{alg:ray_search}
\begin{algorithmic}[1]
\REQUIRE
    Victim model $f$; benign input $\mathbf{x}_b\in\mathbb{R}^n$;\\
    initial direction $\mathbf{d}_{\rm best}\in\{-1,+1\}^n$;\\
    current best radius $r_{\rm best}>0$ (possibly $+\infty$).
\STATE Initialize stage $s=0$ and block index $\ell=1$;
\WHILE{query budget remains}
\STATE $b_s=\min(2^s,n)$;
\STATE $\mathbf{d}_{\rm tmp}=\mathbf{d}_{\rm best}.\operatorname{copy}()$;
\STATE Divide $\mathbf{d}_{\rm tmp}$ into $b_s$ blocks and denote the
       index set of block $\ell$ by $\mathcal{I}_\ell$;
\STATE $\mathbf{d}_{\rm tmp}[\mathcal{I}_\ell]
       =-\mathbf{d}_{\rm tmp}[\mathcal{I}_\ell]$;
\STATE $r_{\rm tmp}=\operatorname{SDBR}
       (f,\mathbf{x}_b,\mathbf{d}_{\rm tmp},r_{\rm best})$;
\IF{$r_{\rm tmp}<r_{\rm best}$}
\STATE $(r_{\rm best},\mathbf{d}_{\rm best})
       =(r_{\rm tmp},\mathbf{d}_{\rm tmp})$;
\ENDIF
\STATE $\ell=\ell+1$;
\IF{$\ell>b_s$}
\STATE $s=s+1$ and $\ell=1$;
\ENDIF
\ENDWHILE
\STATE \textbf{return} $r_{\rm best}$ and $\mathbf{d}_{\rm best}$.
\end{algorithmic}
\end{algorithm}

\begin{figure}[!t]
\centering
\includegraphics[width=0.98\columnwidth]{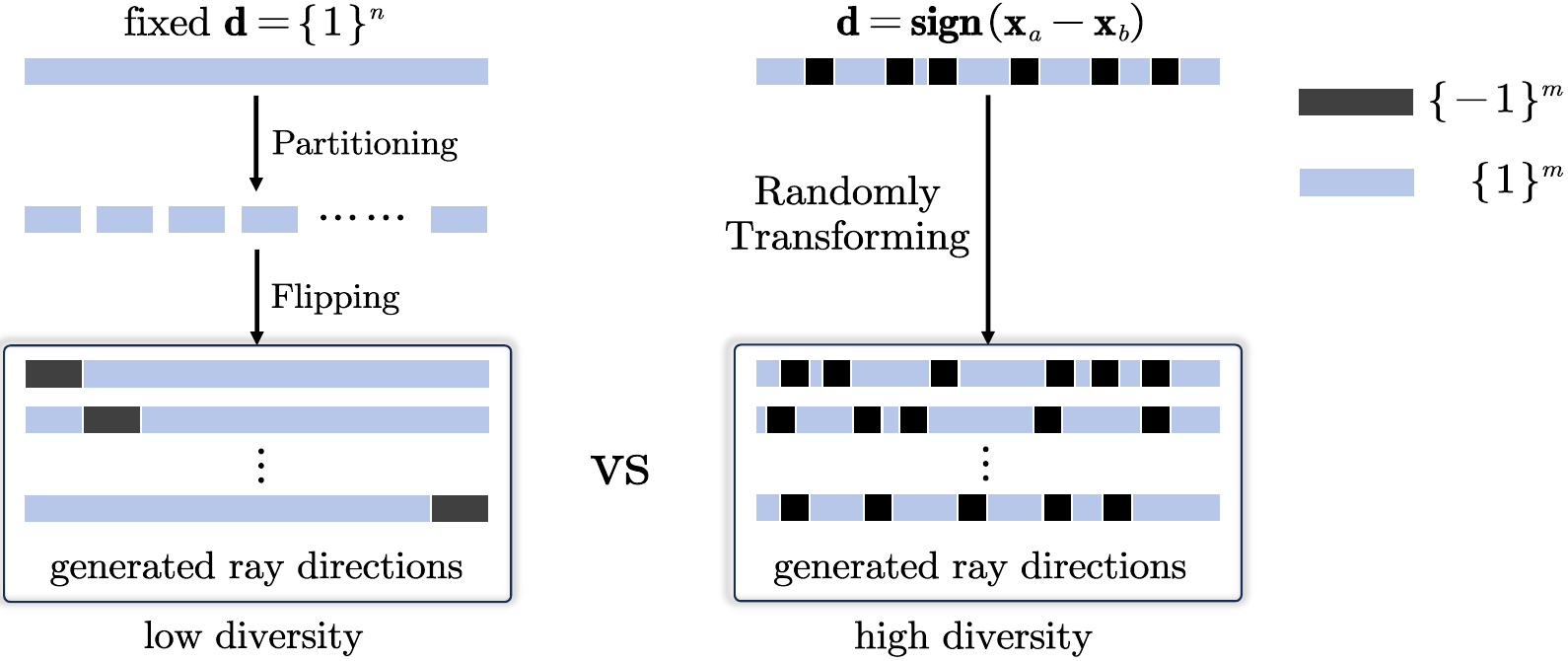}
\caption{Comparison of candidate-direction generation in RayS and GuidedRay.
RayS uses structured block flips from the all-ones direction, whereas
GuidedRay induces varied directions from augmented target-class references.}
\label{fig:sampling_comparison}
\end{figure}

Fig.~\ref{fig:sampling_comparison} shows the difference between the two
initialization mechanisms. RayS flips blocks of the fixed direction
$\{1\}^n$, whereas GuidedRay varies references and augmentations to generate
diverse target-conditioned directions.

\subsection{Untargeted Variant}
\label{subsec:untargeted_variant}
GuidedRay can also be extended to the untargeted setting. Since any class other
than $f(\mathbf{x}_b)$ is acceptable, it requires no target-class reference.
Gaussian noise applied to $\mathbf{x}_b$ produces a candidate whose induced
sign direction is screened by the Fast Test. After success or $N$ failed
tests, the attack respectively applies Ray Search to the accepted direction or
the all-ones direction $\mathbf{d}_{\rm best}=\{1\}^n$.

%% file: sections/experiments.tex
% !TeX spellcheck = en_US
% !TEX root = main.tex

\section{Experiments}
\label{sec:experiments}
%This section performs extensive experiments to verify the effectiveness of GuidedRay.
%Ablation experiments are conducted to verify the motivation behind GuidedRay.

\subsection{Experimental Setup}

\paragraph{Datasets}
We evaluate all attacks on CIFAR-10~\cite{Krizhevsky09},
CIFAR-100~\cite{Krizhevsky09}, and ImageNet~\cite{ILSVRC15}, three widely used
image-classification benchmarks. CIFAR-10 and CIFAR-100 contain $32\times32$
RGB images from 10 and 100 classes, with 6,000 and 600 images per class,
respectively. ImageNet contains 1,000 classes, and its images are resized to
$224\times224$. All pixel values are normalized to $[0,1]$.

For each dataset, we randomly select 1,000 correctly classified images and
conduct one attack trial on each image. In a targeted trial, the target class
$t\neq f(\mathbf{x}_b)$ is selected randomly, and $k\geq1$ correctly classified
samples from class $t$ are used as target-class references. Unless otherwise
specified, targeted attacks are evaluated in the single-reference setting with
$k=1$.

\paragraph{Victim Models}
For the standard setting, we use ResNet-50~\cite{DBLP:conf/eccv/HeZRS16} for CIFAR-10 and CIFAR-100 and
DenseNet-121~\cite{DBLP:conf/cvpr/HuangLMW17} for ImageNet. Their pretrained
weights are obtained from OpenMMLab's MMClassification
toolbox~\cite{2020mmclassification}. The ResNet-50 models achieve top-1
accuracies of 0.952 and 0.799 on CIFAR-10 and CIFAR-100, respectively, while
DenseNet-121 achieves 0.744 on ImageNet.

\paragraph{Attack Performance Metrics}
The \emph{attack success rate} (ASR) at query limit $K$, denoted by ASR@$K$,
is the fraction of trials in which the attack finds, within $K$ queries, an
adversarial example that satisfies both the attack objective and the
$L_\infty$ perturbation threshold $\epsilon$. Query consumption is another
important measure of attack performance. For successful trials, we report the
average and median numbers of queries, denoted by Average Queries (AvQ) and
Median Queries (MeQ), respectively. Since AvQ and MeQ are computed only over
successful trials, we report them together with ASR.

%\begin{figure*}[!htb]
%\centering
%\subfloat[Cifar10+ResNet ($\epsilon = 0.02$)]{\includegraphics[width=0.3\linewidth]{./data/figs/attack_figs/ASR_datasetcifar10_modelresnet50_targeted_norminf_num1000_query10000_epsilon0.02.pdf}}
%\hfil
%\subfloat[Cifar100+ResNet ($\epsilon = 0.02$)]{\includegraphics[width=0.3\linewidth]{./data/figs/attack_figs/ASR_datasetcifar100_modelresnet50_targeted_norminf_num1000_query10000_epsilon0.02.pdf}}
%\hfil
%\subfloat[ImageNet+DenseNet ($\epsilon = 0.1$)]{\includegraphics[width=0.3\linewidth]{./data/figs/attack_figs/ASR_datasetimagenet_modeldensenet121_targeted_norminf_num1000_query10000_epsilon0.1.pdf}}\\
%\vspace{5pt}
%\fbox{\includegraphics[width=0.92\linewidth]{data/figs/attack_figs/legend_standard_targeted.JPG}}
%\caption{ASR against the number of queries for different targeted attack methods on different datasets and models.}
%\label{fig:ASR_standard}
%\end{figure*}

For initialization-stage evaluation, the \emph{initialization success rate}
(ISR) is defined as the fraction of trials in which an adversarial
direction is found within the specified initialization budget.

To summarize the ASR trajectory up to 5,000 queries rather than performance at
a single query limit, we report the normalized area under the ASR--query curve,
denoted by AUC@5K:
\[
\mathrm{AUC@5K}=\frac{1}{5000}\int_{0}^{5000}\mathrm{ASR}(q)\,dq.
\]
We also report $Q@\alpha$, the smallest query count at which the aggregate ASR
over the 1,000 trials reaches $\alpha$. In particular, $Q@20$ and $Q@30$
correspond to ASR levels of $20\%$ and $30\%$, respectively.

\paragraph{Baseline Attacks}
We compare GuidedRay with five representative decision-based attacks:
Bounce~\cite{DBLP:conf/sp/WanFWY24}, HSJA~\cite{DBLP:conf/sp/ChenJW20},
Sign-OPT~\cite{DBLP:conf/iclr/ChengSCC0H20},
Tangent~\cite{DBLP:conf/nips/MaGCYW21}, and
RayS~\cite{DBLP:conf/kdd/ChenG20}. In targeted experiments, GuidedRay and the
first four baselines receive the same target-class reference sample, whereas
RayS starts from the benign input and does not require such a reference.
All evaluated methods operate under the same attacker-resource setting: none
uses a pretrained surrogate model, offline local-model training, or
transfer-based gradient priors. Auxiliary-model-assisted attacks, including
SQBA~\cite{Park2024SQBA}, DEAL~\cite{DBLP:journals/tdsc/ShenLYLZX24}, and
Prior-OPT/Prior-Sign-OPT~\cite{Ma2025PriorRay}, are discussed in
Section~\ref{sec:background} but excluded from the direct comparison because
they require additional resources unavailable to the evaluated attacks.

\paragraph{Defense Mechanisms}
To evaluate GuidedRay against defended models, we further consider
CIFAR-10/ResNet-50 models trained with Adversarial Training
(AT)~\cite{DBLP:conf/iclr/MadryMSTV18} and
TRADES~\cite{DBLP:conf/icml/ZhangYJXGJ19}, using the corresponding checkpoints
released with Tangent Attack~\cite{DBLP:conf/nips/MaGCYW21}.

\paragraph{Augmentation Methods}
In targeted attacks, the candidate-generation module adopts several
augmentation methods, including Gaussian noise, random rotation, random
cropping, and color jitter. Rotation angles are sampled from
$[-180^\circ,180^\circ]$. Random cropping crops an image at a random location,
with the cropping region size ranging from 0.5 to 1.0 of the original image
size, and then resizes the cropped image to the original size. Color
jitter changes brightness, contrast, saturation, and hue. The mixture strategy
samples one of the four methods for each candidate.

\paragraph{Diversity Metric}
We use the average pairwise cosine similarity (APCS) defined in
Eq.~\eqref{eq:apcs_motivation}. A smaller APCS indicates greater diversity
among the generated sign directions. Reported APCS values are averaged across
the attack trials.

%In one attack trial, suppose that there are $m$ directions newly generated for finding an adversarial direction,
%we first compute
%\begin{equation}
%\label{eq:pcs}
%\alpha = \sum_{i=1}^{m}{ 
%\sum_{j=i+1}^{m}{
%\frac{
%2 \mathcal{C} \left( \mathbf{d}_i, \mathbf{d}_j \right)
%}{
%m (m -1)
%}
%}
%} ,
%\end{equation}
%where $\mathbf{d}_i$, $\mathbf{d}_j$ are two directions,
%and $\mathcal{C} (\cdot)$ denotes the cosine similarity function.
%Then, by testing the attack $N_T$ times, we obtain the APCS as follows
%\begin{equation}
%\mathbf{APCS} = \sum_{i=1}^{N_T}{\frac{\alpha _i}{N_T}} ,
%\end{equation}  
%where $\alpha _i$ is the value (as shown in Eq.~\eqref{eq:pcs}) in the $i$-th trial.

\subsection{Targeted Attacks on Undefended Models}
\label{subsec:experiments_undefended}
We first evaluate targeted attacks against undefended victim models. For a
fair comparison, GuidedRay and all reference-based baselines are evaluated in
the single-reference setting, where the attacker is given a benign input and
one target-class reference sample. We then examine how additional target-class
references affect GuidedRay.
Consistent with prior $L_\infty$ decision-based attack evaluations~\cite{
DBLP:conf/kdd/ChenG20,DBLP:conf/nips/MaGCYW21}, the perturbation threshold is
$\epsilon=0.03$ for CIFAR-10 and CIFAR-100 and $\epsilon=0.2$ for ImageNet. In
targeted experiments, we set $N=Q$, allowing
initialization to continue until a feasible direction is found or the total
query budget is exhausted. Each Fast Test counts as one query, and Ray Search
uses only the remaining budget after successful initialization.

\subsubsection{Single-Reference Setting}
Table~\ref{tab:ASR-standard} reports ASR at different query limits. To
highlight the benefit of target-class-guided direction discovery, we also
report the ASR of RayS in the targeted setting. Note that RayS does not require
a target-class reference in its original formulation.

\begin{table}[!htb]
\renewcommand{\arraystretch}{1.10}
\caption{ASR of targeted attacks under the dataset-specific perturbation
thresholds.}
\label{tab:ASR-standard}
\centering
\setlength{\tabcolsep}{5.2pt}
\begin{tabular*}{0.89\columnwidth}{@{\hspace{8pt}\extracolsep{\fill}}cccccc@{\hspace{8pt}}}
\toprule
Dataset /  	&  &  &  &  &	\\
Model /	& Method & @500 & @1K & @3K & @5K\\
$\epsilon$ &  &  &  &  &\\
\midrule
		& Sign-OPT  	&    0.001    &    0.009    &    0.036    &    0.080\\
CIFAR-10 	& HSJA 		&    0.035    &    0.087    &    0.318    &    0.518\\
ResNet-50 	& Tangent 		&    0.016    &    0.091    &    0.318    &    0.510\\
0.03  	& Bounce 		&    0.037    &    0.108    &    0.415    &    0.695\\
		& RayS 		&    0.052    &    0.141    &    0.323    &    0.389\\
\cline{2-6}
		& GuidedRay 	&    \textbf{0.144}    &    \textbf{0.313}    &    \textbf{0.695}    &    \textbf{0.810}\\
\hline
		& Sign-OPT 	&    0.000    &    0.002    &    0.007    &    0.013\\
CIFAR-100	& HSJA 		&    0.004    &    0.014    &    0.065    &    0.166\\
ResNet-50	& Tangent 		&    0.005    &    0.015    &    0.068    &    0.155\\
0.03		& Bounce 		&    0.004    &    0.018    &    0.115    &    0.261\\
		& RayS     		&    0.000    &    0.006    &    0.021    &    0.030\\
\cline{2-6}
		& GuidedRay 	&    \textbf{0.023}    &    \textbf{0.053}    &    \textbf{0.216}    &    \textbf{0.304}\\
\hline
		& Sign-OPT 	&    0.014    &    0.018    &    0.037    &    0.061\\
ImageNet	& HSJA 		&    0.150    &    0.183    &    0.255    &    0.355\\
DenseNet-121	& Tangent 		&    0.147    &    0.167    &    0.271    &    0.333\\
0.2		& Bounce 		&    0.128    &    0.156    &    0.227    &    0.290\\
		& RayS     		&    0.000    &    0.000    &    0.000    &    0.000\\
\cline{2-6}
		& GuidedRay 	&    \textbf{0.171}    &    \textbf{0.246}    &    \textbf{0.356}    &    \textbf{0.402}\\
\bottomrule
\end{tabular*}
\end{table}

Compared with RayS, GuidedRay achieves a pronounced ASR improvement.
At 1,000 queries, GuidedRay improves the ASR over RayS from $0.141$ to $0.313$
on CIFAR-10, from $0.006$ to $0.053$ on CIFAR-100, and from $0$ to $0.246$
on ImageNet.
In Section~\ref{subsubsec:impact_of_aeaf}, 
we demonstrate that the advantage of GuidedRay over RayS is still pronounced, 
even when RayS is allowed to exploit the same target-class reference.

Compared with other state-of-the-art baseline attacks,
GuidedRay outperforms all evaluated baselines in ASR on all three datasets at
every reported query limit from 500 to 5,000. This pattern is consistent with
GuidedRay's principal advantage in
targeted-direction discovery during initialization, before the subsequent
refinement procedures dominate the query cost.

Table~\ref{tab:auc-fixed-asr} compares GuidedRay with the strongest baseline
using AUC@5K and $Q@20$/$Q@30$. AUC is computed over 0--5,000 queries; only
Bounce's $Q@30$ on CIFAR-100 falls beyond this range and is obtained from the
full 10,000-query records.

\begin{table}[!t]
\centering
\renewcommand{\arraystretch}{1.08}
\setlength{\tabcolsep}{2.5pt}
\caption{AUC@5K and queries to reach fixed ASR levels.}
\label{tab:auc-fixed-asr}
\begin{tabular}{llccc}
\toprule
Dataset & Metric & GuidedRay & Best baseline & Rel. gain \\
\midrule
\multirow{3}{*}{CIFAR-10}
& AUC@5K $\uparrow$ & 0.5463 & 0.3381 (Bounce) & 61.6\% \\
& $Q@20$ $\downarrow$ & 640 & 1329 (RayS) & 51.8\% \\
& $Q@30$ $\downarrow$ & 951 & 2133 (Bounce) & 55.4\% \\
\hline
\multirow{3}{*}{CIFAR-100}
& AUC@5K $\uparrow$ & 0.1649 & 0.0972 (Bounce) & 69.7\% \\
& $Q@20$ $\downarrow$ & 2628 & 3990 (Bounce) & 34.1\% \\
& $Q@30$ $\downarrow$ & 4824 & 5679 (Bounce) & 15.1\% \\
\hline
\multirow{3}{*}{ImageNet}
& AUC@5K $\uparrow$ & 0.3080 & 0.2376 (HSJA) & 29.6\% \\
& $Q@20$ $\downarrow$ & 641 & 1557 (HSJA) & 58.8\% \\
& $Q@30$ $\downarrow$ & 1647 & 4043 (HSJA) & 59.3\% \\
\bottomrule
\end{tabular}
\end{table}

GuidedRay improves AUC@5K over the strongest baseline by $61.6\%$, $69.7\%$,
and $29.6\%$ on CIFAR-10, CIFAR-100, and ImageNet, respectively. It also
requires $34.1\%$--$58.8\%$ fewer queries than the fastest baseline to reach
an ASR of $20\%$. The fixed-ASR comparison therefore confirms that the gain is
not tied to one favorable query limit.

\begin{table}[!htb]
\renewcommand{\arraystretch}{1.10}
\caption{Queries required for successful targeted attacks under the
dataset-specific perturbation thresholds and a 5,000-query budget.}
\label{tab:queries_for_succ_attack}
\centering
\begin{tabular}{ccccc}
\toprule
%Dataset (Model) & Method & AvQ & MeQ & ASR \\
Dataset / &   &   &   &   \\
Model / & \multirow{2}{*}{Method} & \multirow{2}{*}{AvQ} & \multirow{2}{*}{MeQ} & \multirow{2}{*}{ASR} \\
$\epsilon$ &   &   &   &   \\
\midrule
		& Sign-OPT  	& 2917.1 & 3135.5 & 0.080 \\
		& HSJA 		& 2556.2 & 2466.0 & 0.518 \\
CIFAR-10 	& Tangent 		& 2543.7 & 2504.5 & 0.510 \\
ResNet-50  	& Bounce		& 2567.9 & 2472.0 & 0.695 \\
0.03      	& RayS 		& 1663.6 & 1302.0 & 0.389 \\
\cline{2-5}
		& GuidedRay 	& \textbf{1628.0} & \textbf{1299.5} & \textbf{0.810} \\
\hline
		& Sign-OPT 	& 2927.0 & 2914.0 & 0.013 \\
		& HSJA 		& 3225.3 & 3566.0 & 0.166 \\
CIFAR-100   & Tangent 		& 3064.0 & 3239.0 & 0.155 \\
ResNet-50      & Bounce 		& 3138.6 & 3192.0 & 0.261 \\
0.03      	& RayS 		& 2288.2 & 2261.0 & 0.030 \\
\cline{2-5}
		& GuidedRay 	& \textbf{2287.0} & \textbf{2188.0} & \textbf{0.304} \\
\hline
		& Sign-OPT 	& 2248.9 & 2146.0 & 0.061 \\
		& HSJA 		& 1653.8 & 994.0 & 0.355 \\
ImageNet    & Tangent 	& 1480.9 & 772.0 & 0.333 \\
DenseNet-121    & Bounce 	& 1451.7 & 749.5 & 0.290 \\
0.2      	& RayS 		& -- & -- & 0.000 \\
\cline{2-5}
		& GuidedRay 	& \textbf{1168.9} & \textbf{644.5} & \textbf{0.402} \\
\bottomrule
\end{tabular}
\end{table}

Table~\ref{tab:queries_for_succ_attack} reports AvQ and MeQ together with
ASR@5K, whose values are consistent with those in Table~\ref{tab:ASR-standard}.
GuidedRay achieves the highest ASR and the lowest AvQ and MeQ across all three
datasets. Compared with RayS, it increases ASR from $38.9\%$ to $81.0\%$ on
CIFAR-10 and from $3.0\%$ to $30.4\%$ on CIFAR-100 while maintaining slightly
lower query consumption. On ImageNet, GuidedRay achieves an ASR of $40.2\%$,
whereas RayS has no successful attack within 5,000 queries.

\subsubsection{Multi-Reference Setting}
\label{subsubsec:k_samples}

As discussed in Section~\ref{subsubsec:k_AEGS}, multiple references provide
additional target-conditioned starting points. We therefore evaluate GuidedRay
with different values of $k$ under a maximum budget of 5,000 queries.
Table~\ref{tab:ASR-pool-num} shows that the overall ASR increases as more
references become available, although the gain is not strictly monotonic at
every individual query limit. The effect is most pronounced on ImageNet, where
ASR@5K increases from $0.402$ for $k=1$ to $0.774$ for $k=30$.

\begin{table}[!t]
\renewcommand{\arraystretch}{1.10}
% \caption{ASR and ISR of GuidedRay-M attacks under different $k$-adversarial examples guided scenarios on different datasets and models.}
\caption{ASR of GuidedRay with different numbers of target-class reference
samples.}
\label{tab:ASR-pool-num}
\centering
\begin{tabular}{cccccc}
\toprule
Dataset /  	&  &  &  &  &	\\
Model /	&	$k$	& @500 & @1K & @3K & @5K\\
$\epsilon$ &  &  &  & &\\
\midrule
CIFAR-10		& 1 		&    0.144 	  &    0.313    &    0.695    &    0.810\\
ResNet-50	& 3 		&    0.171    &    0.348    &    0.746    &    0.891\\
0.03	& 5 		&    0.165    &    0.359    &    0.776    &    0.908\\
	& 10 	&    0.153    &    0.362    &    0.784    &    0.911\\
\hline
		& 1 		&    0.023    &    0.053    &    0.216    &    0.304\\
CIFAR-100	& 3 		&    0.052    &    0.112    &    0.298    &    0.394\\
ResNet-50	& 5 		&    0.040    &    0.102    &    0.276    &    0.408\\
0.03		& 10 	&    0.040    &    0.105    &    0.298    &    0.428\\
		& 50 	&    0.054    &    0.133    &    0.342    &    0.491\\
\hline
		& 1 		&    0.171    &    0.246    &    0.356    &    0.402\\
ImageNet	& 3 		&    0.275    &    0.381    &    0.531    &    0.579\\
DenseNet-121	& 5 		&    0.295    &    0.398    &    0.596    &    0.667\\
0.2		& 10 	&    0.313    &    0.433    &    0.647    &    0.714\\
		& 30 	&    0.362    &    0.509    &    0.713    &    0.774\\
\bottomrule
\end{tabular}
\end{table}

\subsection{Targeted Attacks on Defended Models}
To examine whether GuidedRay retains its advantage in defended scenarios, we
evaluate targeted attacks on CIFAR-10/ResNet-50 models protected by adversarial
training and TRADES. We report ASR at the same four primary query budgets used
for the undefended models: 500, 1,000, 3,000, and 5,000.

\begin{table}[!htb]
\renewcommand{\arraystretch}{1.12}
% \caption{ASR of different targeted attacks against different defense mechanisms by different $\epsilon$ values on the Cifar10 dataset and ResNet model.}
\caption{ASR of targeted attacks against defended CIFAR-10/ResNet-50 models at
$\epsilon=0.03$.}
\label{tab:ASR-defensive}
\centering
\setlength{\tabcolsep}{3.5pt}
\begin{tabular}{llcccc}
\toprule
Defense & Method & @500 & @1K & @3K & @5K\\
\midrule
\multirow{5}{*}{Adv. Training} & Sign-OPT & 0.000 & 0.000 & 0.001 & 0.001\\
& HSJA & 0.009 & 0.011 & 0.011 & 0.012\\
& Tangent & 0.006 & 0.007 & 0.009 & 0.013\\
& Bounce & 0.010 & 0.010 & 0.012 & 0.013\\
& GuidedRay & \textbf{0.019} & \textbf{0.026} & \textbf{0.048} & \textbf{0.057}\\
\hline
\multirow{5}{*}{TRADES} & Sign-OPT & 0.001 & 0.001 & 0.002 & 0.002\\
& HSJA & 0.011 & 0.011 & 0.012 & 0.013\\
& Tangent & 0.005 & 0.005 & 0.007 & 0.007\\
& Bounce & 0.010 & 0.010 & 0.011 & 0.013\\
& GuidedRay & \textbf{0.031} & \textbf{0.036} & \textbf{0.048} & \textbf{0.057}\\
\bottomrule
\end{tabular}
\end{table}

As shown in Table~\ref{tab:ASR-defensive}, GuidedRay achieves the highest ASR
at every reported query limit under both defenses. At 500 queries, it reaches
ASRs of 0.019 under adversarial training and 0.031 under TRADES, compared with
the strongest baseline values of 0.010 and 0.011, respectively. The consistent
advantage across budgets shows that GuidedRay remains effective against both
defended models.

\subsection{Ablation Studies}
\label{subsec:ablation_experiments}
In this subsection, we first evaluate the contribution of ADSF to the attack
performance. We then replace Ray Search with a simpler randomized search
procedure to examine whether the benefit of ADSF remains when the subsequent
optimizer changes. Next, we compare the diversity of the directions generated
during initialization and analyze its relation to initialization success. We
further examine the Fast Test and finally compare the effects of different
candidate augmentation methods.

\subsubsection{Contribution of ADSF}
\label{subsubsec:impact_of_aeaf}
To evaluate the contribution of ADSF to the performance improvement, we
conduct an ablation experiment involving the following three attacks:
\begin{itemize}
\item The first is the original RayS attack.

\item The second is a minimally modified variant of RayS, denoted by RayS-T.
It receives the same target-class reference $\mathbf{x}_a$ as GuidedRay, sets
the starting direction to
$\mathbf{d}_{\rm best}=\operatorname{sign}(\mathbf{x}_a-\mathbf{x}_b)$, and
then applies Ray Search without ADSF.

\item The third is GuidedRay with the complete ADSF initialization.
\end{itemize}
We limit the query budget to 5,000 and evaluate the three attacks on
CIFAR-100/ResNet-50 at $\epsilon=0.03$ in the targeted setting.

\begin{figure}[!htb]
\centering
\includegraphics[width=0.9\linewidth]{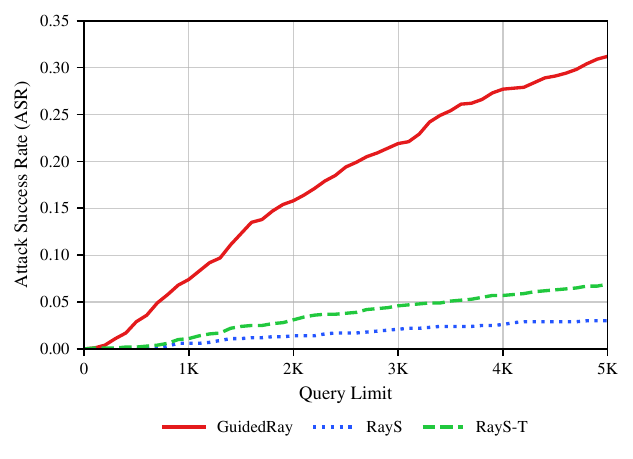}
\caption{ASR of RayS, RayS-T, and GuidedRay on CIFAR-100/ResNet-50 at
$\epsilon=0.03$.}
\label{fig:ASR-method-effectiveness}
\end{figure}

Fig.~\ref{fig:ASR-method-effectiveness} shows that RayS-T provides only a
modest improvement over RayS and still performs substantially worse than
GuidedRay. This comparison shows that the improvement of GuidedRay does not
come solely from replacing the all-ones initialization with one direction
induced by a target-class reference. Generating and screening varied candidate
directions through ADSF provides the larger performance gain.

\subsubsection{Replacing Ray Search}
\label{subsubsec:replace_ray_search}
To separate initialization from the subsequent optimizer, we replace Ray
Search with a simple randomized block-flipping procedure.

Given $\mathbf{d}_{\rm best}\in\{-1,+1\}^n$, the replacement samples an integer
start index $0\leq i_{\rm start}<n$ and block size $1\leq b_s\leq n$, flips coordinates
$i_{\rm start}$ through $\min(i_{\rm start}+b_s-1,n-1)$, and retains the
direction if its decision-boundary radius decreases.

This procedure is simpler than Algorithm~\ref{alg:ray_search}; the resulting
variant is denoted by GuidedRay-W. Table~\ref{tab:ASR-standard-GR-W} compares
it with GuidedRay, RayS, and Bounce on CIFAR-10 and CIFAR-100 at
$\epsilon=0.03$, using the same four primary query budgets as the main
comparison.

\begin{table}[!htb]
\renewcommand{\arraystretch}{1.12}
\caption{ASR after replacing Ray Search with randomized block flipping.}
\label{tab:ASR-standard-GR-W}
\centering
\begin{tabular}{cccccc}
\toprule
Dataset /  	&  &  &  &  &	\\
Model /	& Method & @500 & @1K & @3K & @5K\\
$\epsilon$ &  &  &  & &\\
\midrule
CIFAR-10 	& Bounce 		&    0.037    &    0.108    &    0.415    &    0.695\\
ResNet-50	& RayS 		&    0.052    &    0.141    &    0.323    &    0.389 \\
0.03		& GuidedRay 	&    \textbf{0.144}    &    \textbf{0.313}    &    \textbf{0.695}    &    \textbf{0.810}\\
\cline{2-6}
		& GuidedRay-W &    0.055    &    0.149    &    0.513    &    0.695\\
\hline
CIFAR-100	& Bounce 		&    0.004    &    0.018    &    0.115    &    0.261\\
ResNet-50	& RayS     		&    0.000    &    0.006    &    0.021    &    0.030 \\
0.03		& GuidedRay 	&    \textbf{0.023}    &    \textbf{0.053}    &    \textbf{0.216}    &    \textbf{0.304}\\
\cline{2-6}
		& GuidedRay-W &    0.016    &    0.042    &    0.145    &    0.232\\
\bottomrule
\end{tabular}
\end{table}

GuidedRay-W consistently outperforms RayS at all four evaluated query budgets
and remains competitive with Bounce at earlier budgets. Although replacing Ray
Search reduces the performance of GuidedRay, ADSF remains effective with a
different refinement procedure.

\subsubsection{Candidate-Direction Diversity}
\label{subsubsec:diversity}

We compare the initialization behavior of RayS and GuidedRay. For the diversity
measurement, each method generates 1,000 candidate directions per trial, and
APCS is averaged over 1,000 trials. We then run initialization with a
10,000-query limit and report ISR together with average initialization queries
(Init. AvQ), computed over all trials until an adversarial direction is found
or the initialization budget is exhausted. GuidedRay is evaluated with $k=1$
and $k=10$, whereas RayS uses its original reference-free initialization.

\begin{table}[!htb]
\renewcommand{\arraystretch}{1.12}
\caption{Diversity and success of the initialization phase. Lower APCS and
initialization AvQ, together with higher ISR, indicate better performance.
The number of target-class references $k$ applies only to GuidedRay.}
\label{tab:diversity-comparison}
\centering
\setlength{\tabcolsep}{5pt}
\begin{tabularx}{0.9\linewidth}{@{}YYYYY@{}}
\toprule
\multicolumn{5}{c}{CIFAR-10 / ResNet-50}   \\
\hline
Attack          & $k$   &   ISR         & Init. AvQ  & APCS \\
\hline
RayS            & --    & 0.416         & 5848.36    &    0.994 \\
GuidedRay     & 1		& 0.906 		& 1292.10    &    0.464  \\
GuidedRay		& 10		& 0.985 		& 333.75     &    0.301  \\
\toprule
\multicolumn{5}{c}{CIFAR-100 / ResNet-50}   \\
\hline
Attack          & $k$   &  ISR          & Init. AvQ  & APCS \\
\hline
RayS            & --    & 0.135         & 8974.67    &    0.994 \\
GuidedRay     & 1		& 0.412 		& 6275.92    &    0.506  \\
GuidedRay		& 10	    & 0.613 		& 4538.60    &    0.339  \\
\toprule
\multicolumn{5}{c}{ImageNet / DenseNet-121}   \\
\hline
Attack          & $k$   &   ISR         & Init. AvQ  & APCS \\
\hline
RayS            & --    & 0.000         & 10000.00   &    0.961 \\
GuidedRay     & 1		& 0.420 		& 6264.25    &    0.449  \\
GuidedRay		& 10	    & 0.751 		& 3289.42    &    0.343  \\
\bottomrule
\end{tabularx}
\end{table}

\paragraph{Initialization comparison}
%Table~\ref{tab:diversity-comparison} shows partial experimental results,
%highlighting the advantage of the initialization phase of GuidedRay.
%It is clear that the APCS corresponding to RayS almost equals to the upper bound $1$,
%demonstrating that the diversity of newly generated ray directions 
%in the initialization phase of RayS is extremely low.
%On the contrary, the APCS corresponding to GuidedRay is small.
%At the same time, the ISR (resp. AvQ) of RayS is low (resp. high), even approaches 0 (resp. 10,000).
%As a comparison, the ISR (resp. AvQ) of GuidedRay is high (resp. low).
%Particularly, in the case of Cifar10/ResNet, 
%compared to RayS, the ISR of GuidedRay increases from $0.416$ to $0.985$ when $k=10$. 
%And AvQ decreases from $5848.36$ to $333.75$, which is improved by a facor of about $17.5$.

Table~\ref{tab:diversity-comparison} shows that GuidedRay produces
substantially lower APCS than RayS and simultaneously achieves higher ISR with
fewer initialization queries. Increasing $k$ from 1 to 10 further lowers APCS,
increases ISR, and reduces initialization AvQ on all three datasets. For
example, on CIFAR-10/ResNet-50, GuidedRay with $k=10$ increases ISR from
$0.416$ to $0.985$ relative to RayS and reduces initialization AvQ from
$5848.36$ to $333.75$, a factor of approximately $17.5$.

\begin{figure*}[!htb]
\centering
\includegraphics[width=\linewidth]{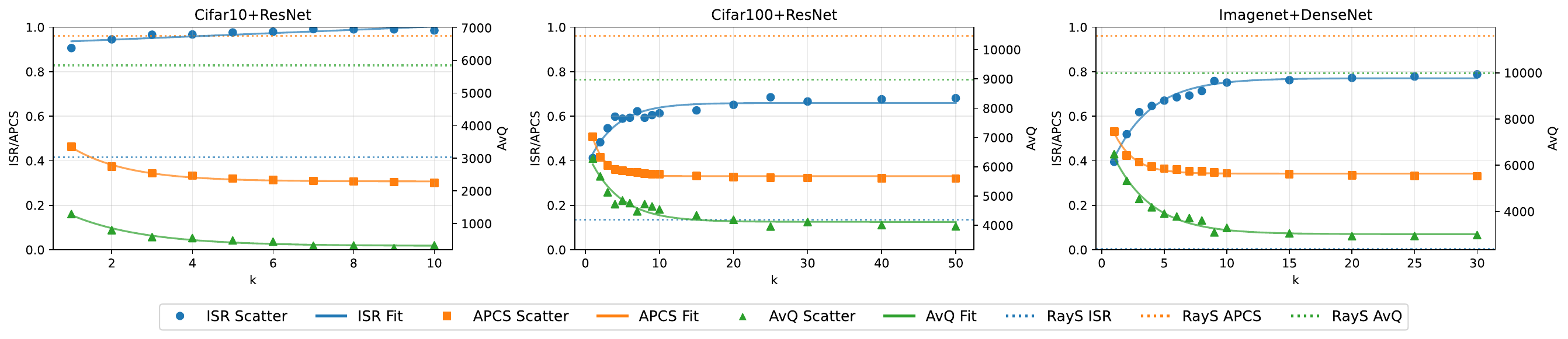}
\caption{APCS, ISR, and initialization AvQ of GuidedRay with different numbers
of target-class references.}
\label{fig:apcs-avq-isr-k-GR}
\end{figure*}

\begin{figure*}[!htb]
\centering
\subfloat[CIFAR-10/ResNet-50]{\includegraphics[width=0.32\linewidth]{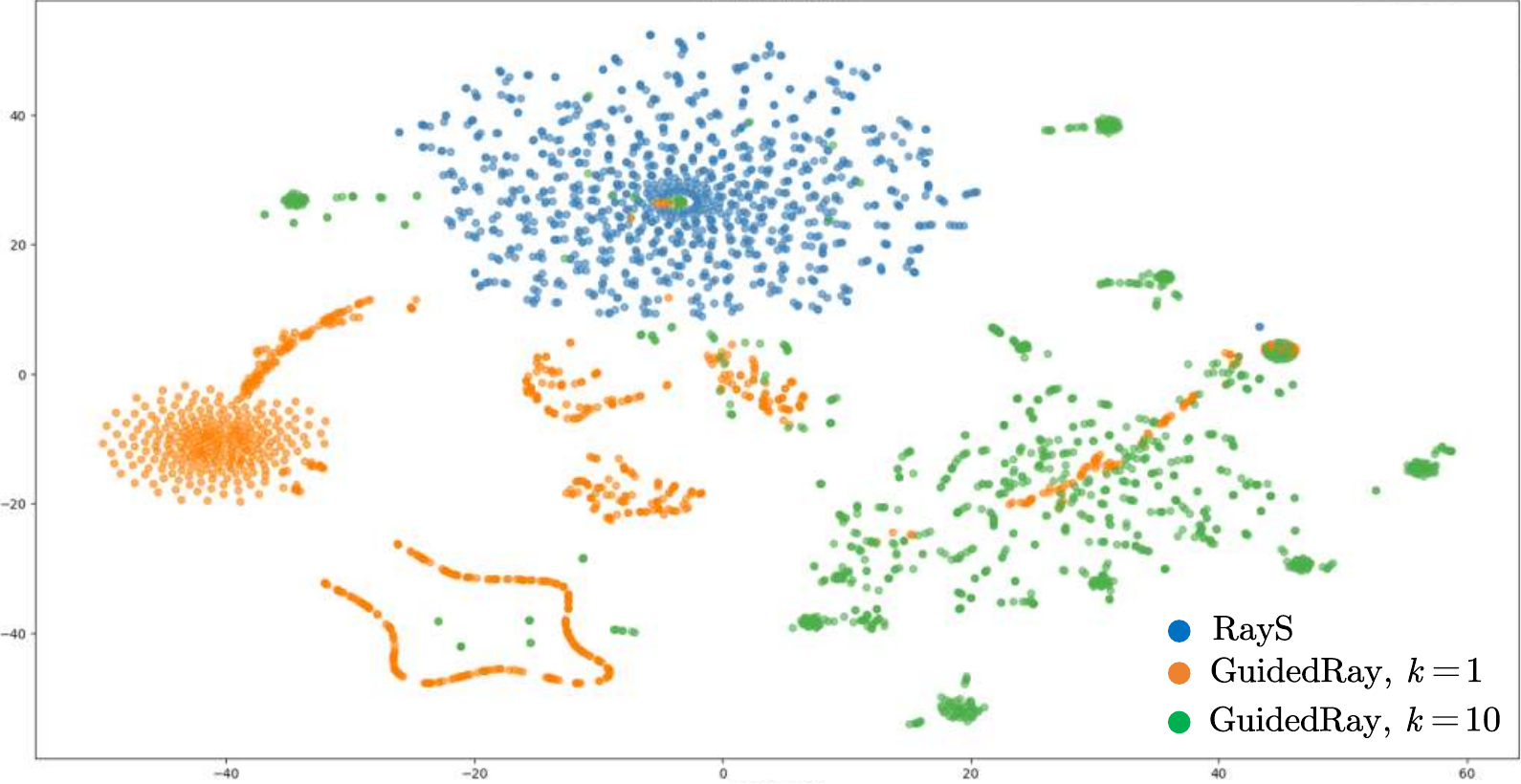}}
\hfil
\subfloat[CIFAR-100/ResNet-50]{\includegraphics[width=0.32\linewidth]{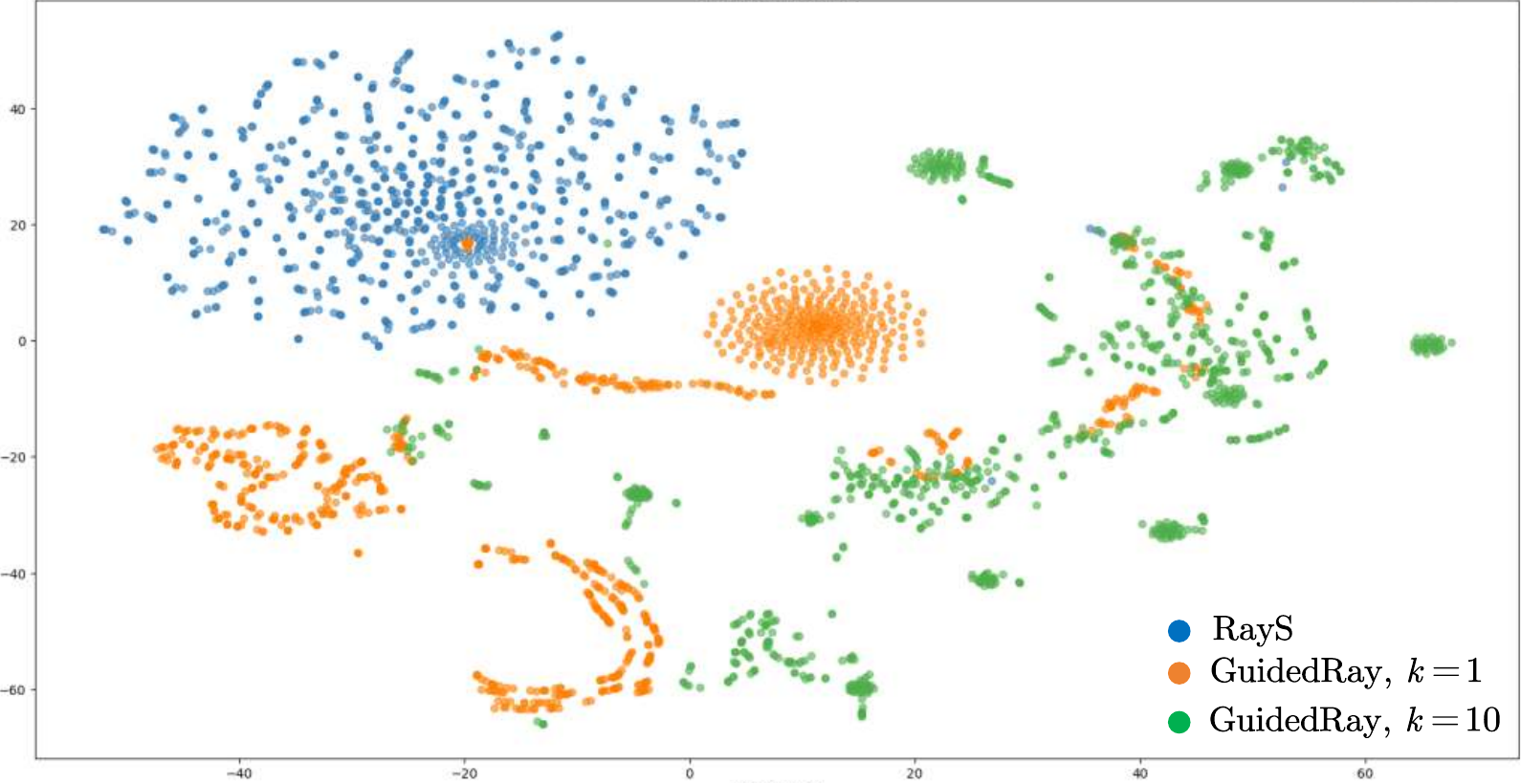}}
\hfil
\subfloat[ImageNet/DenseNet-121]{\includegraphics[width=0.32\linewidth]{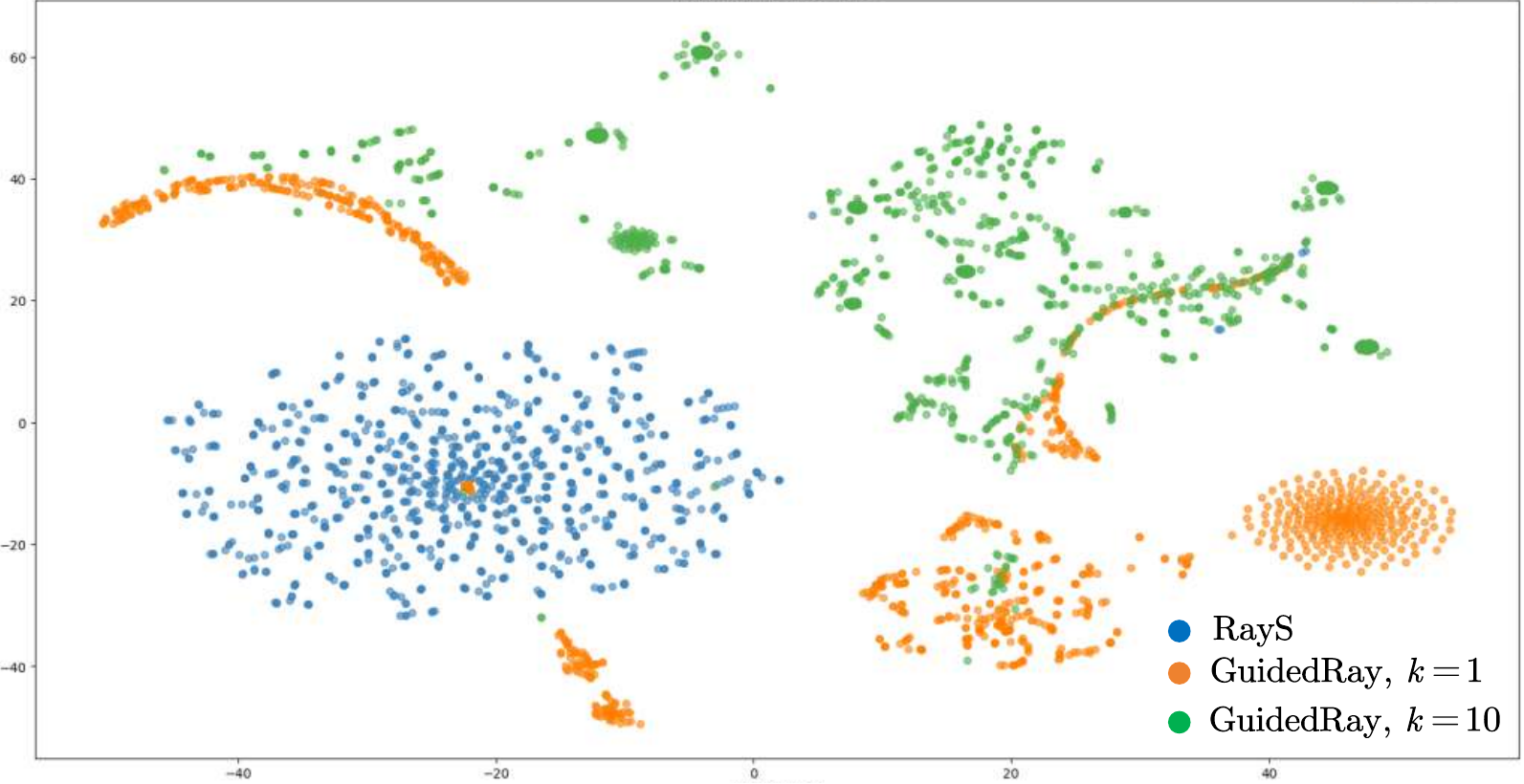}}\\
\caption{t-SNE visualization of the generated sign directions.}
\label{fig:diversity_show}
\end{figure*}

\paragraph{Relation between Diversity and Initialization}
Fig.~\ref{fig:apcs-avq-isr-k-GR} reports APCS, ISR, and initialization AvQ
under different settings of $k$. Increasing $k$ is generally associated with
lower APCS and initialization AvQ and higher ISR. The scatter plots and fitted
curves summarize these trends, with almost all fits achieving a coefficient of
determination ($R^2$) above $0.9$.

\paragraph{Direction visualization}
Using t-SNE~\cite{maaten2008visualizing}, Fig.~\ref{fig:diversity_show}
visualizes directions collected from one trial for the same benign input and
target class. GuidedRay covers a broader region than RayS, and increasing $k$
further spreads the generated directions. The broader spread visually
supports the lower APCS values of GuidedRay. Together, the APCS results and
t-SNE visualization show that GuidedRay generates more diverse candidate
directions than RayS.

\subsubsection{Fast-Test Acceptance Rate}
\label{subsubsec:analysis_of_test}

For each dataset, we sample 5,000 pairs consisting of a correctly classified
benign input $\mathbf{x}_b$ and a target-class reference $\mathbf{x}_a$,
construct $\mathbf{d}=\operatorname{sign}(\mathbf{x}_a-\mathbf{x}_b)$, and
apply the Fast Test. The acceptance rates are $42.74\%$, $37.9\%$, and
$47.2\%$ on CIFAR-10, CIFAR-100, and ImageNet, respectively. Because a rejected
direction may still be feasible at another radius, these values quantify the
fraction of reference-induced directions certified by the one-query test.
Despite requiring only one model query, the Fast Test certifies a substantial
fraction of reference-induced directions, allowing ADSF to avoid costly
boundary searches for candidates that fail the screening test.

\subsubsection{Effect of Candidate Augmentations}
\begin{figure*}[!t]
\centering
\subfloat[CIFAR-10 ($\epsilon=0.03$)]{\includegraphics[width=0.30\linewidth]{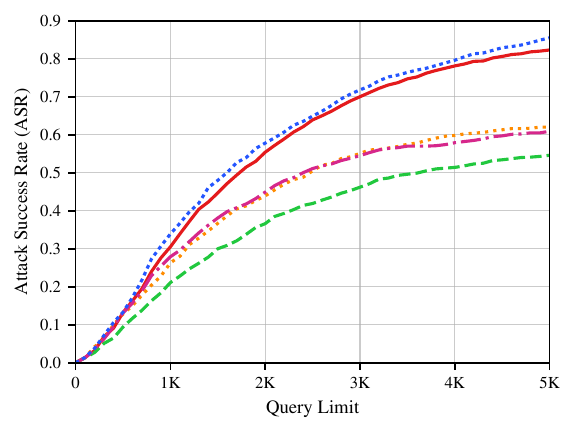}}
\hfil
\subfloat[CIFAR-100 ($\epsilon=0.03$)]{\includegraphics[width=0.30\linewidth]{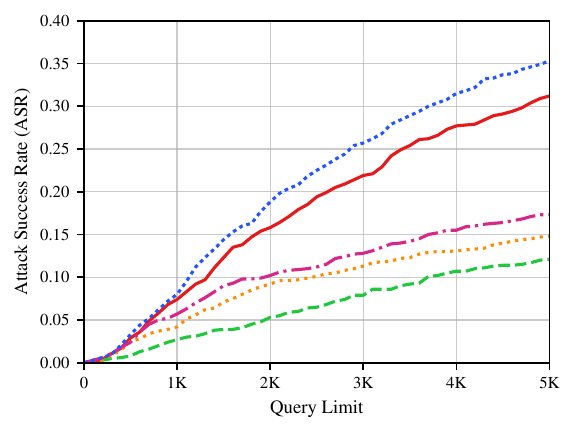}}
\hfil
\subfloat[ImageNet ($\epsilon=0.2$)]{\includegraphics[width=0.30\linewidth]{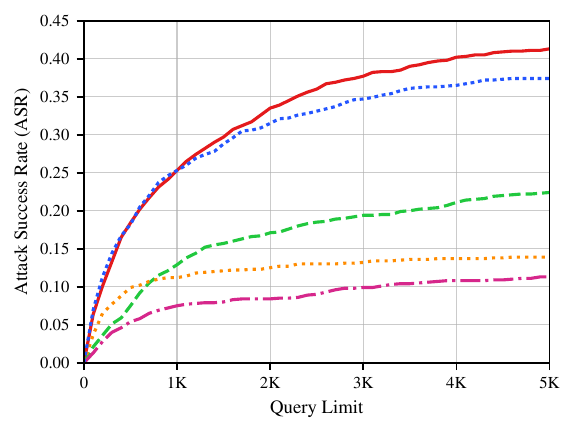}}\\
\vspace{5pt}
\fbox{\includegraphics[width=0.92\linewidth]{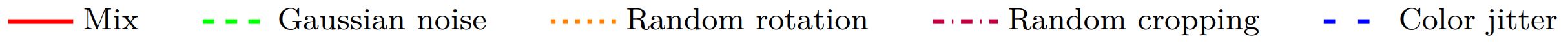}}
\caption{ASR curves for different candidate augmentation methods.}
\label{fig:ASR-augment-method}
\end{figure*}

We compare Gaussian noise, random rotation, random cropping, color jitter, and
their mixture under a 5,000-query budget while keeping the remaining attack
settings unchanged. As shown in Fig.~\ref{fig:ASR-augment-method}, color jitter
is the strongest individual augmentation across all three datasets, while the
mixture remains competitive on CIFAR-10 and CIFAR-100 and achieves the highest
ASR on ImageNet. Since the effectiveness of individual augmentations varies
across datasets, we adopt the mixture as the default to avoid dataset-specific
augmentation selection.

\subsection{Untargeted Attacks}
GuidedRay can also be applied to untargeted attacks through the variant
described in Section~\ref{subsec:untargeted_variant}. The perturbation threshold is
$\epsilon=0.01$ for CIFAR-10 and CIFAR-100 and $\epsilon=0.03$ for ImageNet.
We report results up to 5,000 queries, and GuidedRay tests at most $N=50$
candidate directions during initialization.

As shown in Table~\ref{tab:ASR-standard-untargeted}, GuidedRay achieves higher ASR
than RayS at every reported query limit on CIFAR-10 and CIFAR-100. At 1,000
queries, for example, ASR increases from $0.230$ to $0.285$ on CIFAR-10 and
from $0.453$ to $0.495$ on CIFAR-100. These results indicate that its
initialization strategy remains useful even when the attack objective is
untargeted. The gains are nevertheless smaller than those observed in the
targeted setting, where identifying a direction toward one prescribed class is
substantially more restrictive.

\begin{table}[H]
\renewcommand{\arraystretch}{1.12}
\caption{ASR of untargeted attacks under the dataset-specific perturbation
thresholds.}
\label{tab:ASR-standard-untargeted}
\centering
\setlength{\tabcolsep}{5.2pt}
\begin{tabular*}{0.89\columnwidth}{@{\hspace{8pt}\extracolsep{\fill}}cccccc@{\hspace{8pt}}}
\toprule
Dataset /  	&  &  &  &  &	\\
Model /	& Method & @500 & @1K & @3K & @5K\\
$\epsilon$ &  &  &  &  &\\
\midrule
	& Sign-OPT 	&    0.012    &    0.019    &    0.042    &    0.061\\
CIFAR-10	& HSJA 		&    0.075    &    0.158    &    0.300    &    0.397\\
ResNet-50	& Tangent 		&    0.074    &    0.149    &    0.288    &    0.396\\
0.01		& Bounce 		&    0.080    &    0.188    &    0.363    &    0.463\\
		& RayS 		&    0.107    &    0.230    &    0.427    &    0.541\\
\cline{2-6}
		& GuidedRay 	&    \textbf{0.169}    &    \textbf{0.285}    &    \textbf{0.474}    &    \textbf{0.566}\\
\hline
		& Sign-OPT 	&    0.040    &    0.058    &    0.100    &    0.131\\
CIFAR-100	& HSJA 		&    0.153    &    0.274    &    0.428    &    0.520\\
ResNet-50	& Tangent 		&    0.171    &    0.269    &    0.428    &    0.518\\
0.01		& Bounce 		&    0.161    &    0.304    &    0.468    &    0.546\\
		& RayS 		&    0.283    &    0.453    &    0.664    &    0.743\\
\cline{2-6}
		& GuidedRay 	&    \textbf{0.366}    &    \textbf{0.495}    &    \textbf{0.684}    &    \textbf{0.758}\\
\hline
		& Sign-OPT 	&    0.047    &    0.049    &    0.070    &    0.078\\
ImageNet	& HSJA 		&    0.103    &    0.126    &    0.160    &    0.192\\
DenseNet-121	& Tangent 		&    0.102    &    0.117    &    0.156    &    0.183\\
0.03		& Bounce 		&    0.101    &    0.125    &    0.168    &    0.201\\
		& RayS 		&    \textbf{0.479}    &    \textbf{0.676}    &    \textbf{0.884}    &    0.934\\
\cline{2-6}
		& GuidedRay 	&    0.467    &    \textbf{0.676}    &    0.880    &    \textbf{0.939}\\
\bottomrule
\end{tabular*}
\end{table}

On ImageNet, GuidedRay and RayS perform similarly: their ASRs differ by at most
$0.012$ across the four query limits and are equal at 1,000 queries.
This smaller performance gap is
consistent with the less restrictive untargeted objective, for which reaching
any class other than the benign class is sufficient. Overall, the results show
that GuidedRay remains effective in the untargeted setting, while its more
pronounced benefits arise in targeted attacks that require a direction to
reach one prescribed class.

% \begin{figure*}[!t]
% \centering
% \subfloat[Cifar10+ResNet ($\epsilon = 0.01$)]{\includegraphics[width=0.32\linewidth]{data/figs/attack_figs/ASR_datasetcifar10_modelresnet50_untargeted_norminf_num1000_query10000_epsilon0.01.pdf}}
% \hfil
% \subfloat[Cifar100+ResNet ($\epsilon = 0.01$)]{\includegraphics[width=0.32\linewidth]{data/figs/attack_figs/ASR_datasetcifar100_modelresnet50_untargeted_norminf_num1000_query10000_epsilon0.01.pdf}}
% \hfil
% \subfloat[ImageNet+DenseNet ($\epsilon = 0.03$)]{\includegraphics[width=0.32\linewidth]{data/figs/attack_figs/ASR_datasetimagenet_modeldensenet121_untargeted_norminf_num1000_query10000_epsilon0.03.pdf}}\\
% \vspace{5pt}
% \fbox{\includegraphics[width=0.95\linewidth]{data/figs/attack_figs/legend_standard_untargeted.JPG}}
% \caption{ASR against the number of queries for different untargeted attack methods on different datasets and models.}
% \label{fig:ASR_standard-untargeted}
% \end{figure*}

%\subsection{Generalizability of Diversity-Oriented Sampling}